\documentclass[final,leqno]{siamltex}

\usepackage{graphicx}
\usepackage{dcolumn}

\usepackage{amsmath,amssymb,amstext,amscd,mathrsfs,eucal,bm}
\usepackage{subfigure}
\usepackage{cases}
\usepackage{epic,eepic,graphics,pspicture,overpic}

\usepackage{tabularx}

\usepackage{mathtools}
\usepackage{color}
\usepackage{tikz}
\usepackage{tkz-euclide} 
\usetkzobj{angles} 

\newcounter{subeq}

\DeclareMathAlphabet\mathbfcal{OMS}{cmsy}{b}{n}
\renewcommand{\theenumi}{\roman{enumi}}

		\def\beq{\begin{equation}}
		\def\eeq{\end{equation}}
		\newcommand{\eref}[1]{(\ref{eqn:#1})}
		\newcommand{\elab}[1]{\label{eqn:#1}}
		
		\newcommand{\fref}[1]{\ref{fig:#1}}
		\newcommand{\flab}[1]{\label{fig:#1}}
		
		\newcommand{\sref}[1]{\ref{sec:#1}}
  		\newcommand{\slab}[1]{\label{sec:#1}}

		\newcommand{\av}[1]{\overline{#1}}
		\newcommand{\nor}[1]{\lVert #1 \rVert}
		\newcommand{\dpar}[2]{\frac{\partial #1}{\partial #2}}
		\newcommand{\dt}[2]{\frac{\mathrm{d} #1}{\mathrm{d} #2}}

		\begingroup
		\renewcommand*{\arraystretch}{1.5}
		\endgroup
	
			\newcommand*\colvec[3][]{
			    \renewcommand*{\arraystretch}{.7}\begin{pmatrix}\ifx\relax#1\relax\else#1\\\fi#2\\#3\end{pmatrix}
			}

\def\dd{\mathrm{d}}

\def\ee{\mathrm{e}}

\def\bx{\boldsymbol{x}}

\def\bu{\boldsymbol{u}}

\def\H{\mathcal{H}}
\def\L{\mathcal{L}}
\def\I{\mathcal{I}}

\def\bphi{\boldsymbol{\phi}}
\def\b0{\boldsymbol{0}}

\def\bvarphi{\boldsymbol{\varphi}}

\def\Pe{\text{Pe}}
\def\Da{\text{Da}}

\title{FKPP fronts in cellular flows: \\ the large-P\'eclet regime}

\author{Alexandra Tzella\thanks{School of Mathematics, University of Birmingham, Edgbaston, Birmingham, B15 2TT, United Kingdom (a.tzella@bham.ac.uk).}
        \and Jacques Vanneste\thanks{School of Mathematics and Maxwell Institute for Mathematical Sciences, University of Edinburgh, King's Buildings,
  Edinburgh EH9 3FD, United Kingdom (J.Vanneste@ed.ac.uk).}}

\begin{document}

\maketitle

\begin{abstract}
We investigate the propagation of chemical fronts arising in Fisher--Kolmogorov--Petrovskii--Piskunov (FKPP) type models in the presence of a steady cellular flow. In the long-time limit, a steadily propagating pulsating front is established. Its speed, on which we focus, can be obtained by solving an eigenvalue problem closely related to large-deviation theory.  We employ asymptotic methods to solve this eigenvalue problem in the limit of small molecular diffusivity (large P\'eclet number, $\Pe \gg 1$) and arbitrary reaction rate (arbitrary Damk\"ohler number $\Da$).

 We identify three regimes corresponding to the distinguished limits $\Da = O(\Pe^{-1})$, $\Da=O\left((\log \Pe)^{-1}\right)$ and $\Da = O(\Pe)$ and, in each regime, obtain the front speed  in terms of a different non-trivial function of the relevant combination of $\Pe$ and $\Da$. Closed-form expressions for the speed, characterised by power-law and logarithmic dependences on $\Da$ and $\Pe$ and valid in intermediate regimes, are deduced as limiting cases. 
Taken together, our asymptotic results provide a complete description of the complex dependence of the front speed on $\Da$ for $\Pe \gg 1$. They are confirmed by numerical solutions of the eigenvalue problem determining the front speed, and illustrated by a number of numerical simulations of the advection--diffusion--reaction equation.
\end{abstract}



\pagestyle{myheadings}
\thispagestyle{plain}
\markboth{A. TZELLA AND J. VANNESTE}{FKPP FRONTS IN CELLULAR FLOWS}

\section{Introduction}
In a wide variety of  environmental and engineering applications, 
chemical or 
 biological reactions in fluids propagate  
in the form of localized, strongly inhomogeneous structures associated with reactive fronts \cite{Tel_etal2005,NeufeldHernandezGarcia2009}.  
These are usually established as a result of the interaction between  molecular diffusion, local growth and saturation, but their propagation can be greatly facilitated by advection by a flow.
There has been a growing interest in analysing this impact of advection on the propagation of reactive fronts, as indicated by the large number of experimental,  and theoretical studies (e.g., \cite{PocheauHarambat2008,SchwartzSolomon2008,Thompson_etal2010,BargteilSolomon2012}) and \cite{Xin2000,Berestycki2003,Xin2000b}, respectively).

Much of this work focusses on the effect of incompressible two-dimensional periodic flows, and in particular on the cellular vortex flow. Introduced by  \cite{Childress1979}, 
 this is a steady flow  
with streamfunction
\beq\elab{psiA}
\psi(x,y)=-U\sin (x/\ell)\sin (y/\ell),
\eeq
where $U$ is the maximum flow speed and  $2\pi\ell$ is the period in both $x$ and $y$. 
 When the system is confined between parallel, impermeable  walls at $y=0$ and $\pi\ell$, as considered in this paper, the flow consists of a one-dimensional infinite array of vortices rotating in alternating directions (see Fig.\ \fref{streamfnarray}).  
These vortices are confined within cells that are bounded by a separatrix  connecting  a network of hyperbolic stagnation points.

\begin{figure*}
	\centerline{\includegraphics[width=\linewidth]{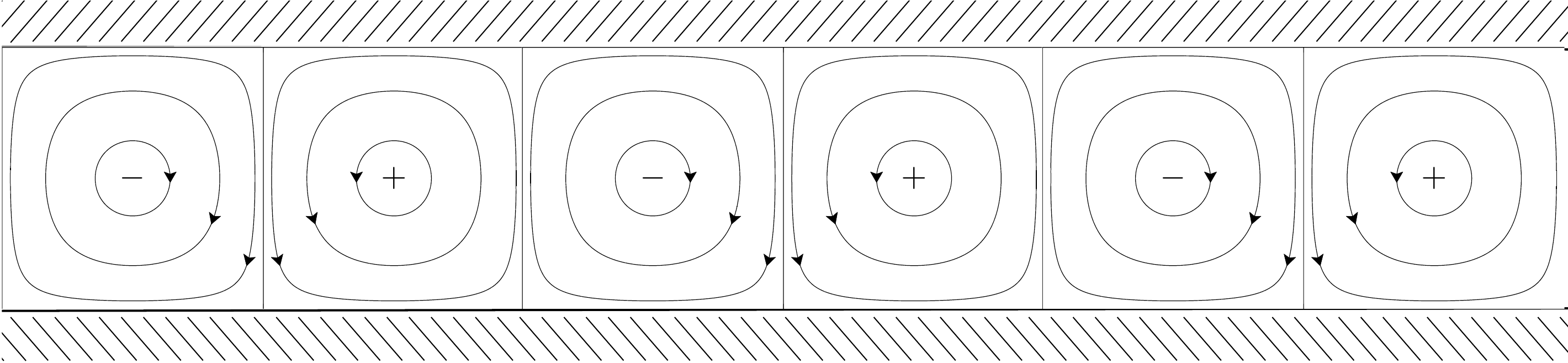}}
\caption{
 Schematic of the streamlines for the cellular vortex  flow with streamfunction \eref{psiA}, confined between  parallel, impermeable walls. 
 The half-cells with anticlockwise (clockwise) circulation are denoted by the $+$ ($-$) signs.  
 }
\flab{streamfnarray}
\end{figure*}

In the absence of advection, the simplest model of front propagation is the FKPP model, named after the pioneering work by Fisher \cite{Fisher1937} and Kolmogorov, Petrovskii and Piskunov \cite{Kolmogorov_etal1937}. This model describes the evolution of a single constituent that diffuses and undergoes a logistic growth, leading to the formation of a steadily travelling front. 
In the presence of a cellular flow (or more general steady periodic flows), the corresponding advection--diffusion--reaction model admits pulsating front solutions that change periodically with respect to time as they travel \cite{BerestyckiHamel2002}. 

The behaviour of these pulsating fronts depends on  two non-dimensional parameters: the 
Damk\"ohler and P\'eclet   numbers, 
\[
\Da=\ell/(U \tau)
\quad\text{and}\quad 
\Pe=U\ell/\kappa,
\]
where $\tau$ is the reaction time and 
$\kappa$  the molecular diffusivity, which measure
the  strength of advection  relative to reaction 
and to diffusion, respectively.
In the interpretation of our results, we will consider a fixed geometry and a fixed  flow, in which case the values of $\Pe$ and $\Da$ are controlled by  
$\kappa$ and $\tau$, respectively. In practice, however, it is easier to achieve this control by varying  $\ell$ and $U$
(see e.g. \cite{PocheauHarambat2008}).

 This paper focusses on the limit of large $\Pe$, relevant to many applications where advection dominates over diffusion. This is a singular limit, of course,  since the weak diffusion leads to the creation of spatial scales that are vanishingly small as $\Pe \to \infty$. 
These small scales are apparent in  
Figure \fref{fronts} which illustrates the dependence of the front structure on the reaction time by showing  snapshots of the concentration for different
 Damk\"ohler number $\Da$ at fixed (large) $\Pe=250$. For small $\Da$ (slow reaction, Fig. \fref{fronts}(a)), the front spreads across several cells, with high concentrations within boundary layers surrounding the separatrix. For intermediate $\Da$ (Fig. \fref{fronts}(b)), the front is narrower: its leading edge is confined around the separatrix as it invades successive cells. For large $\Da$ (fast reaction, Fig. \fref{fronts}(c)), the front is very sharp with a leading edge that penetrates into the cell interiors.

\begin{figure*}
		\centerline{(a) Regime I} 
	\centerline{\includegraphics[width=\linewidth]{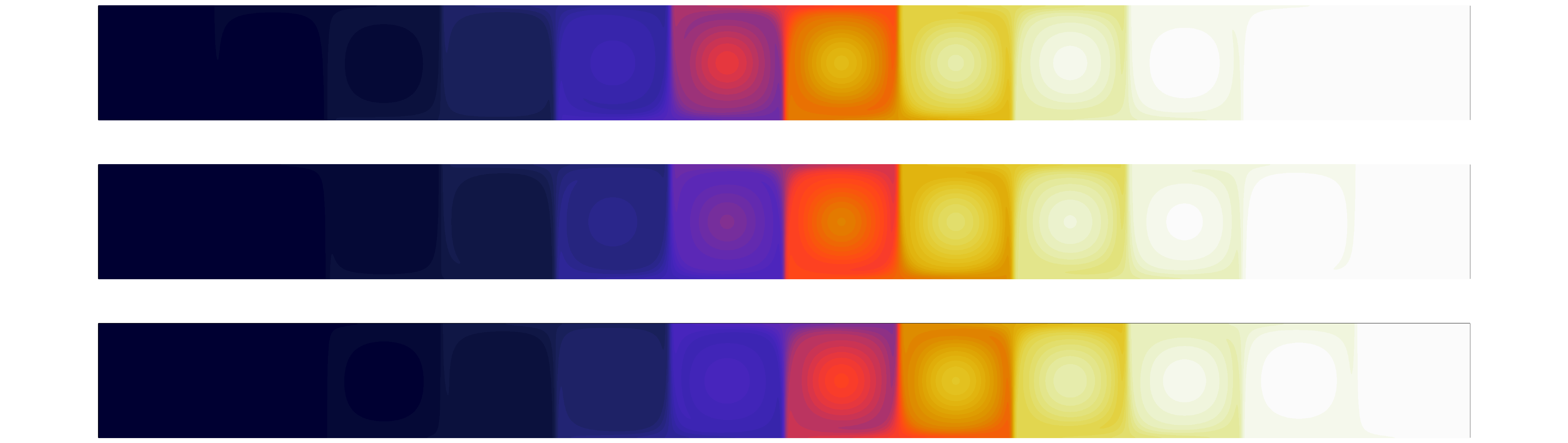}}
	\vspace{.4cm}
	\centerline{(b) Regime II} 
	\centerline{\includegraphics[width=\linewidth]{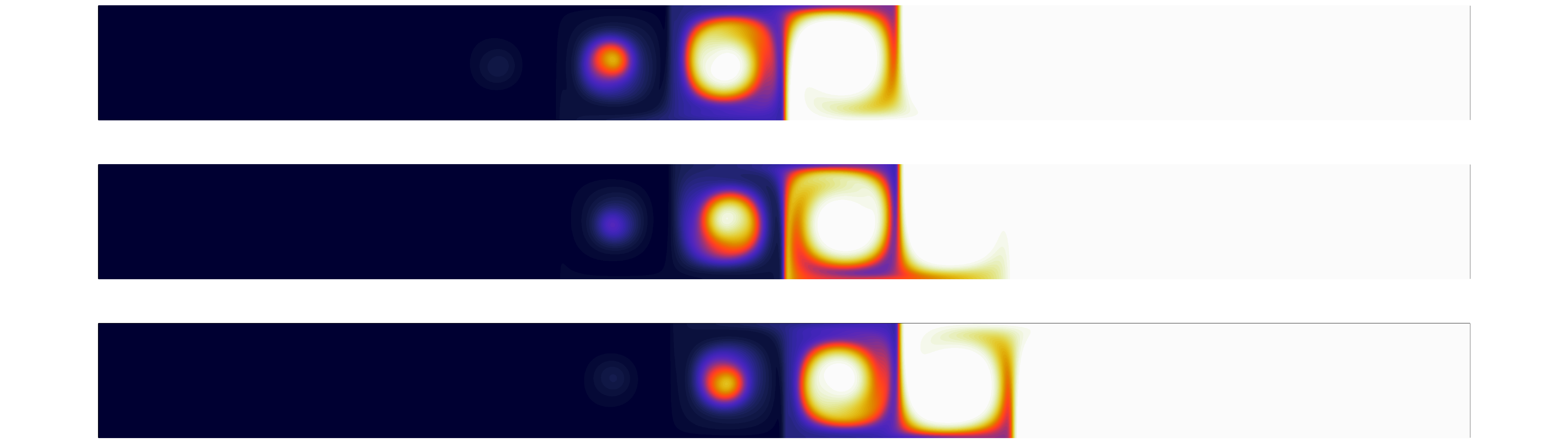}}
		\vspace{.4cm}
	\centerline{(c) Regime III}  
	\centerline{\includegraphics[width=1.04\linewidth]{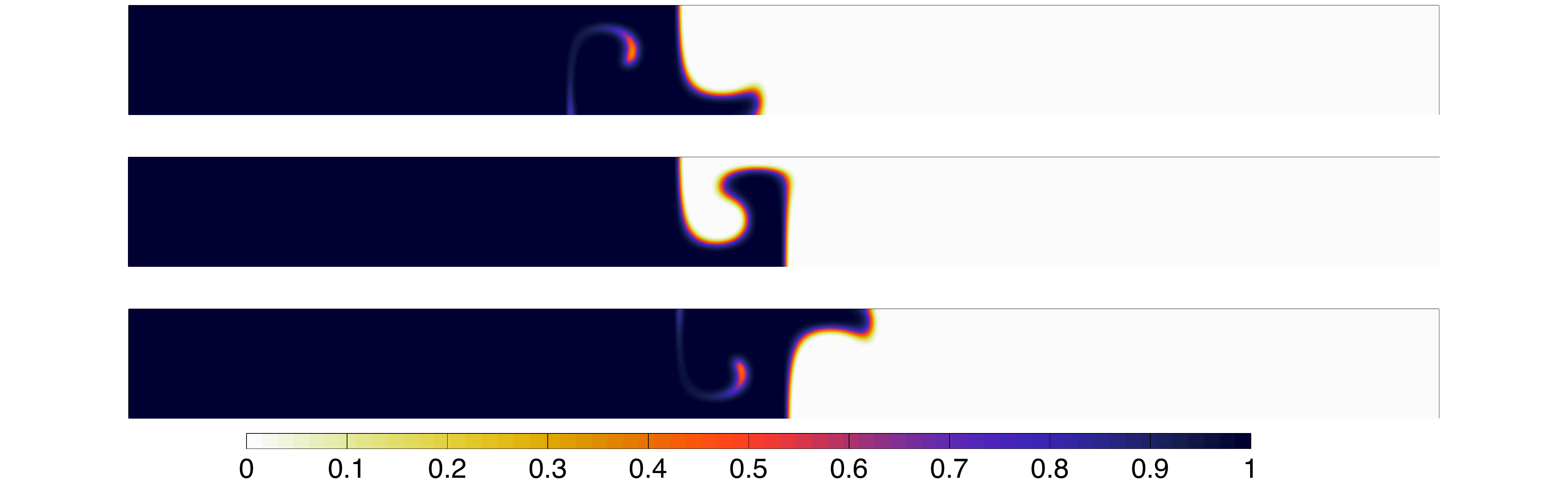}}
		\vspace{.4cm}
\caption{
(Color online). Snapshots of the concentration $\theta$ for $\Pe=250$, illustrating the different type of fronts depending on the reaction rate. 
The reaction rate is determined by:  (a) $\Da=4\times 10^{-2}$ (corresponding to a front speed $c\approx 0.15$), (b) $\Da=4\times 10^{-1}$ ($c\approx 0.44$) and (c) $\Da=4$ ($c\approx 0.67$). In each case,  three successive  snapshots separated by a time interval $\pi\ell/(2c)$ are shown.
  }
\flab{fronts}
\end{figure*}

The main quantitative characteristic of the front is its long-time speed, $c$. This speed is a function of $\Da$ and $\Pe$ only when the initial conditions are sufficiently close to a step function. Assuming this, Freidlin and G\"artner \cite{GertnerFreidlin1979} showed that $c$ can be deduced from the principal eigenvalue of a certain linear operator. This eigenvalue can be interpreted in the framework of large-deviation theory: specifically, it is the Legendre dual of the  rate function $g(c)$ associated with the probability density function for the position of fluid particles that have been displaced -- by advection and diffusion -- to a  distance $ct$ in a time $t\gg 1$. 
Intuitively, these particles control the concentration near the leading edge of the front which, by linearisation, is
approximately of the form  
$\exp(-t(g(x/t)-\Da))$, whence the front speed  $c=x/t=g^{-1}(\Da)$ is obtained. 
An alternative approach, based on the minimum speed of propagation, leads to the same eigenvalue problem, as established in   \cite{Weinberger2002,Berestycki_etal2005}.

The eigenvalue problem does not provide an explicit analytical expression for the front speed but
needs to be solved numerically, through computations that become increasingly intensive as  $\Pe\to \infty$ or $\Da \to \infty$. In the present paper, we carry out a detailed asymptotic analysis of the eigenvalue problem for $\Pe \gg 1$ and arbitrary $\Da$. This provides simpler, and in some cases completely explicit, expressions for the front speed, extracting the dominant scalings and elucidating the physical mechanisms of propagation depending on the relative values of $\Pe$ and $\Da$.

Partial results of this type have been derived for slow reaction i.e.\ for $\Da=O(\Pe^{-1})$:
 the dimensionless front speed 
was argued to scale like  $c/U =O(\Pe^{-3/4})$ in \cite{Audoly_etal2000}.   
This scaling prediction is in agreement with rigorous bounds obtained in \cite{NovikovRyzhik2007}  and was confirmed by numerical simulations     \cite{Abel_etal2001,Abel_etal2002,Vladimirova_etal2003}.  
It  is consistent with the closed-form prediction obtained using a homogenization technique  which is however only valid   for $\Da\ll\Pe^{-1}$.  
 In this regime, $c$ is found to be proportional to the square root of the effective diffusivity deduced from a linear cell problem \cite{Constantin_etal2001,Heinze_etal2001,Heinze2005,RyzhikZlatos2007,Zlatos2010} and  determined in \cite{Soward1987,Shraiman1987,Rosenbluth_etal1987}.  In the opposite limit of fast reaction, i.e.\ for $\Da=O(\Pe)$, 
 $c$ can be deduced from the homogenization of a Hamilton--Jacobi equation  \cite{Freidlin1985,MajdaSouganidis1994,FreidlinSowers1999} and computed by  minimizing a certain action functional \cite{TzellaVanneste2014}. The present paper extends these results to provide a complete description of the asymptotics of $c$ as $\Pe \to \infty$.

\section{Main results and outline}\slab{results}
\begin{table}
\caption{The three distinguished scalings of $\Da$  appearing in the asymptotics of the front speed $c$ for $\Pe \gg 1$. The scalings are  associated with three regimes that correspond to the three  types of fronts depicted in Figure \fref{fronts}. 
In each regime, the speed of the front is expressed in terms of a non-trivial function $\mathscr{C}_i, i=1,\,2,\,3$, that involves a distinct combination of $\Pe$ and $\Da$. The range of validity of each expression is also indicated.}
\begin{center} \footnotesize
\begin{tabular}{cccc} 
Regime & $\Da$ &  c/$U$  & Range of validity  \\[2 pt] 
\hline
\\[1 pt] 
I            & $O(\Pe^{-1})$                      & $\Pe^{-{3}/{4}}\mathscr{C}_{1}(\Pe\Da)$ &  $\Da \ll (\log \Pe)^{-1}$
\\[10pt]
II           & $O((\log\Pe)^{-1})$      & $(\log\Pe)^{-1}\mathscr{C}_{2}(\Da\log\Pe)$ & $\Pe^{-1} \ll \Da \ll \Pe$
\\[10pt]
III          & $O(\Pe)$                 &  $\mathscr{C}_{3}(\Da/\Pe)$ & $\Da \gg (\log \Pe)^{-1}$
\\[10pt] 
\hline
\\
\end{tabular}
\end{center} 
\label{tbl1}
\end{table}

We carry out an asymptotic analysis of the eigenvalue problem determining $c$ and identify three distinguished regimes, characterised by the value of  $\Da$ relative to $\Pe$. These three regimes  correspond to  the three  types of fronts depicted in Figure \fref{fronts}. 
In each regime, we obtain  the front speed in terms of a non-trivial function of a combination of $\Pe$ and $\Da$ (see Table \ref{tbl1}). The function relevant to each regime is obtained by solving one-dimensional problems numerically. We moreover show that the three regimes overlap for intermediate values of $\Da$, thus confirming that our results cover the whole range of  $\Da$.

Our derivation of $c$ in the first two regimes exploits the matched-asymptotics analysis recently carried out by Haynes and Vanneste \cite{HaynesVanneste2014b}. Their paper considers the dispersion of particles in an unbounded cellular flow and derives the rate function $g$ from which we infer $c$ (after some adaptation to account for the walls). 
The analysis captures the behaviour of the concentration in the interior of the cells at the leading edge of the front. In Regime I, the concentration is found to be nearly constant along  
the streamlines (see  Fig.\ \fref{fronts}(a)), while in Regime II the concentration is vanishing inside the cell's interior (see  Fig.\ \fref{fronts}(b)). In both regimes, a boundary layer around the separatrix is crucial for the front dynamics.
In Regime III, where the reaction is fast, we rely on a Wentzel--Kramers--Brillouin--Jeffreys (WKB) approach which shows that $c$ is controlled by a single action-minimising trajectory \cite{TzellaVanneste2014}.

We note that our predictions are formal, involving no rigorous estimates of the associated errors.   
Instead, they are verified against  values of $c$ derived from  (i) numerical solutions of  the principal eigenvalue problem, and (ii) direct numerical simulations of the
FKPP advection--diffusion--reaction equation.  
Figure \fref{c} shows that  the asymptotic expressions for $c$ are in  excellent agreement with the corresponding values obtained from the eigenvalue problem. 

\begin{figure*}
\begin{center}
\begin{minipage}{\linewidth}
\begin{minipage}{0.57\linewidth}
\begin{center}	
\begin{overpic}[width=1\linewidth]{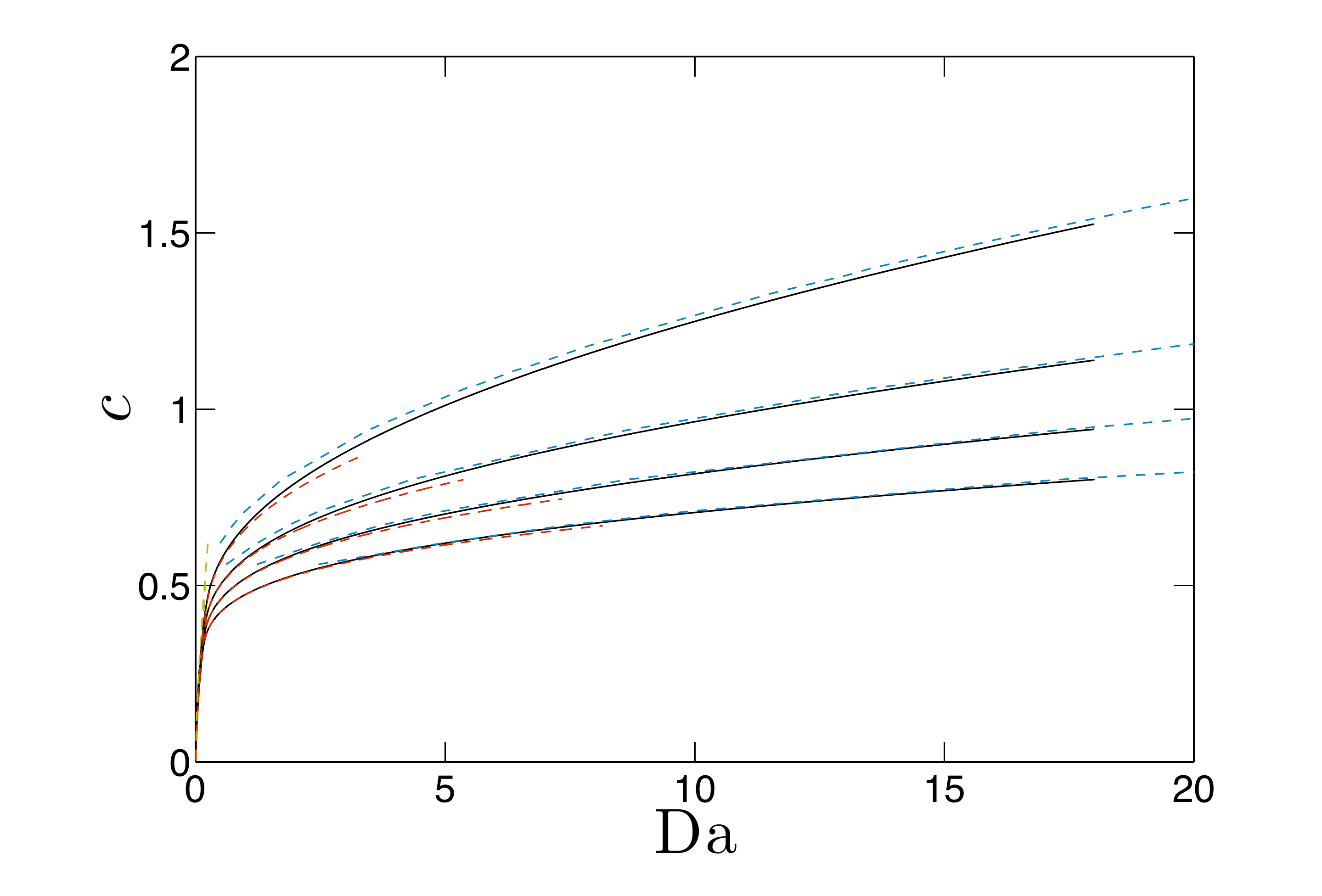}
\put(80,59){\rotatebox{12}{\scriptsize\textcolor{black}{$\Pe=50$}}}
\put(80,47){\rotatebox{9}{\scriptsize\textcolor{black}{$\Pe=125$}}}
\put(80,41.5){\rotatebox{7}{\scriptsize\textcolor{black}{$\Pe=250$}}}
\put(80,37){\rotatebox{6}{\scriptsize\textcolor{black}{$\Pe=500$}}}
\end{overpic}	
\end{center}	
\centerline{(a)}
\end{minipage}
\begin{minipage}{0.42\linewidth}
\centerline{\includegraphics[width=1\linewidth]{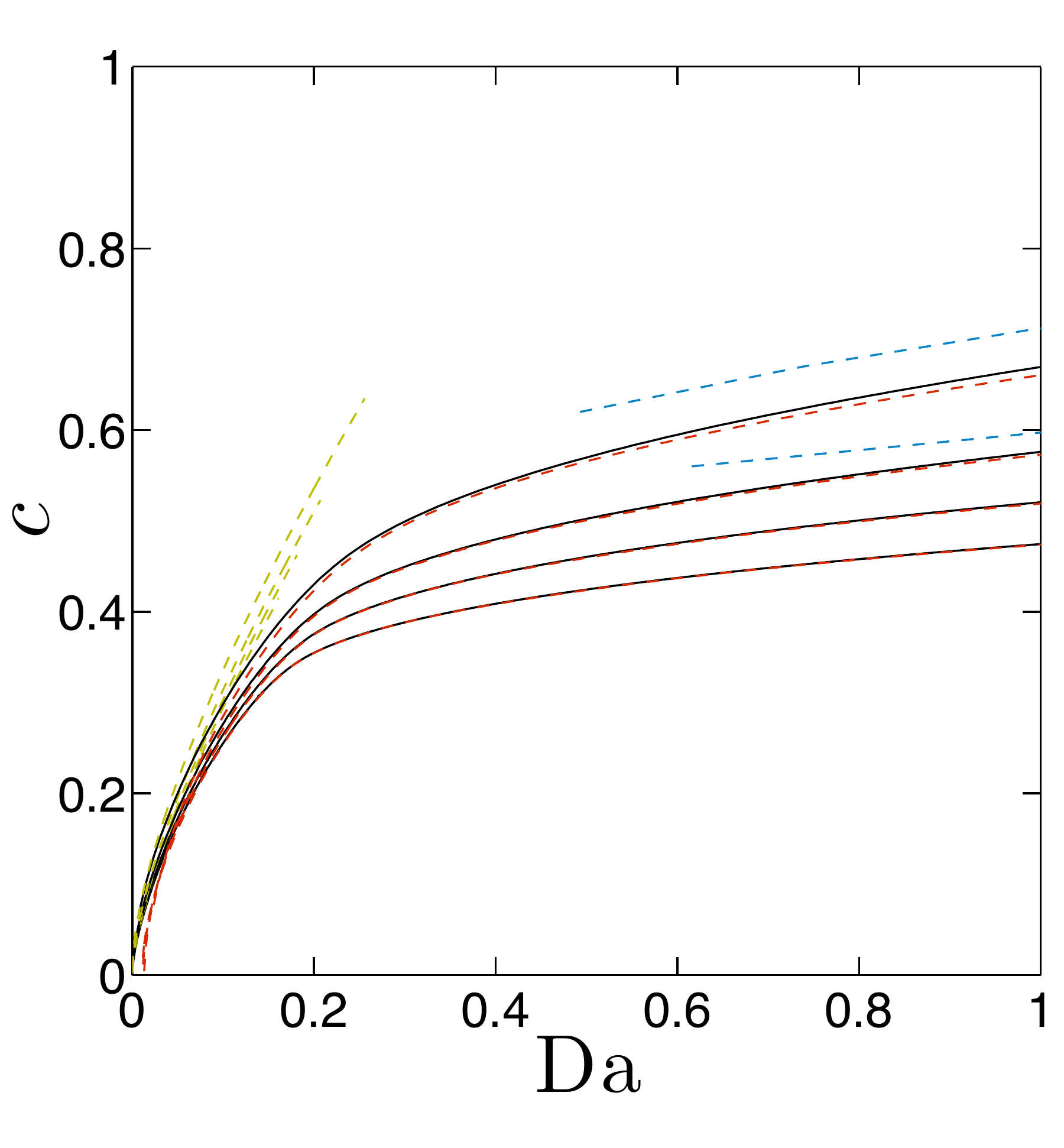}}
\centerline{(b)}
\end{minipage}
\end{minipage}
\end{center}
\caption{(Color online). (a) Comparison between  numerical and asymptotic predictions for the front speed $c$ as a function of $\Da$ and for different values of $\Pe$. The numerical results (solid, black lines) are obtained from \eref{speed} by solving the eigenvalue problem \eref{eval2} numerically. The asymptotic results (colored, dashed lines) correspond to three distinguished regimes describing the three types of fronts shown in Figure \fref{fronts} with the associated predictions reported in Table \ref{tbl1}. (b) Same as (a) but focussing on small values of $\Da$.  
}
\flab{c}
\end{figure*}

There are four subregimes in which the asymptotic expressions  for $c$ reduce to  closed forms.  The reduced expressions, which in fact cover  most of the $(\Da,\Pe)$-plane for $\Pe \gg 1$,
are summarised in Table   \ref{tbl2}. They are used to verify the overlap between regimes mentioned above. In these subregimes, $c$ behaves qualitatively as follows.
For $\Da\ll\Pe^{-1}$, the diffusive approximation obtained from classical homogenisation theory is recovered. For $\Pe^{-1} \ll \Da \ll (\log \Pe)^{-1}$, $c$ is proportional to $\Da^{3/4} (\log \Pe)^{-1/4}$ and is controlled by the dynamics along the separatrix, with the hyperbolic stagnation points at the cell corners playing a negligible role. 
The range $(\log\Pe)^{-1}\ll\Da\ll\Pe$ captures the slow growth  of $c$ with $\Da$  
which, in contrast, can be attributed to the stagnation points. 
The expression for  $c$ in this range can in fact be crudely approximated as the $\Da$-independent $c \sim \pi/\log\Pe$. This is qualitatively similar to the expression   obtained in  
 \cite{Abel_etal2002,Cencini_etal2003}  using a heuristic approach  based on 
  an alternative model, the so-called G-equation 
  (see also \cite{Nolen_etal2009} for a more rigorous analysis). 
     To our knowledge, no equivalent expression  has previously been derived from the eigenvalue problem.  
For $\Da\gg\Pe$, the reaction is so strong that advection contributes only a small correction to the well-known FKPP speed 
$c_0 = 2U\sqrt{\Da/\Pe}=2\sqrt{\kappa/\tau}$.

The paper is structured as follows. In section \ref{sec:formulation}, we give a brief derivation of the eigenvalue problem for the front speed $c$. The relation between the eigenvalue problem and large-deviation theory is also described there. 
Sections \sref{RegimeI}, \sref{RegimeII} and \sref{RegimeIII} are devoted to each of the three distinguished asymptotic regimes. The explicit expressions for $c$ in the four subregimes reported in Table \ref{tbl2} are also derived in these sections.
Comparisons with numerical results are presented in section \sref{num}.
The paper ends with the concluding section \sref{conc}. Technical details are relegated to three Appendices.  
A word of caution about our notation may be necessary: to avoid a proliferation of symbols, we use the same letters to denote quantities that are scaled differently in each of the three Regimes. Specifically, we systematically denote by $f_0$ and $\hat q$ the (leading-order) Legendre duals to $g$ and $c$ suitably scaled in each Regime, and by $\gamma$ the combination of $\Da$ and $\Pe$ on which $c$ depends transcendentally (this is the argument of each of the functions $\mathscr{C}_{i}$ in Table \ref{tbl1}). This should not lead to confusion since these scaled quantities are used exclusively and independently in each of the sections \sref{RegimeI}--\sref{RegimeIII}.

\begin{table}
\caption{The four subregimes and the corresponding closed-form  expressions for the front speed. Here $\nu\approx 0.53$ ,and $\text{W}_\mathrm{p}$ denotes the principal real branch of the Lambert W function \cite{NIST:DLMF}. 
}
\begin{center} \footnotesize
\begin{tabular}{ccc} 
Subregime  & Range of validity & c/$U$  \\[.3ex]
 \hline
\\[.3ex]
Ia & $\Da\ll\Pe^{-1}$ & $\displaystyle (8\nu)^{1/2}\Pe^{-1/4}\Da^{1/2}$ \\[10pt]
Ib/IIa & $\Pe^{-1}\ll\Da\ll(\log\Pe)^{-1}$ &  $\displaystyle\pi\nu^{1/2}(4/3)^{3/4}\Da^{3/4}(\log\Pe)^{-1/4}$ \\[10pt]
IIb/IIIa & $\displaystyle(\log \Pe)^{-1}\ll\Da\ll\Pe$ & $\displaystyle\pi/\text{W}_\mathrm{p}(8\Pe/\Da)$ \\[10pt]
IIIb & $\Da\gg\Pe$ & $\displaystyle 2\sqrt{\Da/\Pe}(1+3\Pe/(16\Da))$\\[10pt]
\hline
\end{tabular}
\label{tbl2}
\end{center}
\end{table}

\section{Eigenvalue problem for the front speed}\slab{formulation}

We investigate the propagation of a reactive front  that is established in the cellular flow with streamfunction \eref{psiA}. 
The governing equation is the FKPP advection--diffusion--reaction equation that describes the evolution of the reactive concentration $\theta(\bx,t)$. 
Taking $\ell$ as reference length and the advective time scale $\ell/U$ as reference time, this equation takes the non-dimensional form 
\beq\elab{FKPP}
 \partial_t\theta+\bu\cdot\nabla\theta=\Pe^{-1}\Delta\theta+
\Da\,r(\theta),
\eeq 
where 
$\bm{u}=(u_1,u_2)=(-\partial_y \psi,\partial_x\psi)$   
and 
\beq\elab{psi}
\psi(x,y)=-\sin x\sin y,
\eeq
are the dimensionless velocity and streamfunction.
Here, the reaction term is $r(\theta)=\theta(1-\theta)$ or, more generally, any function $r(\theta)$ that 
satisfies 
$r(0)=r(1)=0$
with 
 $r(\theta)>0$ for $\theta \in(0,1)$, $r(\theta)<0$ for $\theta \notin[0,1]$ and $r'(0)=\sup_{0<\theta<1} r(\theta)/\theta=1$.
 We take the domain to be an infinite two-dimensional strip with no-flux boundary conditions
 \beq\elab{noflux}
 \partial_y\theta= 0 \quad\text{at} \ \ y=0, \, \pi, 
\eeq
 and $\theta\to 1$ as $x\to-\infty$, $\theta\to 0$ as $x\to\infty$, 
 so that the front advances rightwards. 
As initial condition we take   
$\theta(x,y,0)=\Theta(-x)$,
where $\Theta$ is the Heaviside step function. Note that our non-dimensionalisation implies that the front speed $c$ will from now on be expressed relative to the flow velocity $U$, as reported in Tables \ref{tbl1} and \ref{tbl2}.

G\"artner and Friedlin \cite{GertnerFreidlin1979} showed that the long-time speed of propagation of the front can be determined by the behaviour of the solution near the front's leading edge.
There, $\theta\ll 1$ and $r(\theta)\approx r'(0)\theta=\theta$ so that
equation \eref{FKPP} becomes
\beq\elab{lin}
\partial_t\theta+\bu\cdot\nabla\theta=\Pe^{-1}\Delta\theta+\Da\,\theta.
\eeq
For $t \gg 1$, the solution can be written as the multiscale expansion
\beq\elab{expansion}
  \theta(\bx,t)=t^{-1/2}\ee^{t(\Da-g(\xi))}\left(\phi_0(\bx,\xi)+t^{-1}\phi_1(\bx,\xi)+\cdots \right),
\eeq  
where  
\beq\elab{c}
   \xi={x}/{t}=O(1),
\eeq
 is treated as a slow parameter. The $\Pe$-dependent function $g(\xi)$ is independent of $\Da$ and characterises the dispersion of purely passive particles. It can be recognised as the rate (or Cram\'er) function
of large-deviation theory, which quantifies the rough asymptotics of the probability density function of the particle positions for $t \gg 1$ \cite{HaynesVanneste2014a}. The functions
 $\phi_i, \, i=0,1, 2,\cdots,$ are periodic in $x$:  $\phi_i(x+2\pi,y)=\phi_i(x,y)$. The boundary conditions \eref{noflux} further imply that 
 \beq\elab{phi0b}
 \partial_y\phi_i= 0 \quad\text{at} \ \ y=0,\, \pi.
 \eeq

Substituting \eref{expansion} into \eref{lin} and equating powers of $t^{-1}$ yields, at leading order,  an eigenvalue problem for $\phi_0$. Dropping the subscript $0$ for convenience, this reads
 \beq\elab{eval2}
 \Pe^{-1}\Delta\phi-\bu\cdot\nabla\phi
-2{\Pe}^{-1}q\partial_x\phi+\left(u_1q+\Pe^{-1}q^2\right)\phi
=f(q)\phi, 
  \eeq
where  $q=g'(\xi)$ can be treated as a parameter and $f(q)=\xi g'(\xi)-g(\xi)$ is the eigenvalue. 
 The relevant eigenvalue is the principal eigenvalue (that with maximum real part) because   
it corresponds to the  slowest decaying solution of \eref{expansion}.
The Krein--Rutman theorem implies that this eigenvalue is unique, real and isolated, with a positive associated eigenfunction $\phi>0$. 
Moreover, $f(q)\geq 0$ and is convex  \cite{BerestyckiHamel2002}, so that 
 $f(q)$ and $g(c)$ are related by  a Legendre transform  
\beq\elab{leg}
 g(\xi)=\sup_{q}(q\,\xi-f(q))\quad\text{and}\quad f(q)=\sup_{\xi}(q\,\xi-g(\xi)).
\eeq

With $g(\xi)$ determined, the front speed  may be obtained heuristically by observing that the solution to \eref{lin} must neither grow nor decay exponentially with time in a reference frame moving with the front, i.e., for $\xi=x/t=c$. This happens precisely when $g(c)=\Da$ which suggests that the front speed satisfies
\beq\elab{speed}
 c=g^{-1}(\Da).
\eeq
(Note that subdominant terms in expansion \eref{expansion} do not influence the above expression for the long-time speed value.)
The rigorous treatment in  \cite{GertnerFreidlin1979} confirms this to be the correct speed. 
An alternative argument seeks solution to \eref{lin} of the form $\exp(-qx+(f(q)+\Da)t)\phi$, recovering the eigenvalue problem \eref{eval2}. The front speed is then determined from the minimum speed condition
\beq\elab{speed2}
c=\inf_{q>0}\frac{f(q)+\Da}{q},
\eeq
first introduced in \cite{GertnerFreidlin1979} 
and easily checked to be equivalent to \eref{speed} (see also Ch.\ 7 in \cite{Freidlin1985},  \cite{Evans_etal1989} and \cite{Weinberger2002,Berestycki_etal2005}). In what follows, we rely on the form \eref{speed} of the front speed: this makes direct contact with recent large-deviation results obtained in \cite{HaynesVanneste2014a,HaynesVanneste2014b} for the problem of a non-reacting passive scalar (i.e., $\Da=0$) in an unbounded cellular flow which we use in our treatment of Regimes I and II.

The   eigenvalue problem  \eref{eval2} -- in fact a family of eigenvalue problems paramerized by $q$ -- plays a central role in this paper.  
In the absence of  flow, $f(q)=q^2/\Pe$, recovering the classical formula for the speed $c_0=2\sqrt{\Da/\Pe}=2\sqrt{\kappa/\tau}$.
For general $\bu\neq\bm{0}$, the  eigenvalue problem \eref{eval2} cannot be solved analytically. 
Numerically, it can be obtained  by straightforward discretisation. 
 Computations  are simplified by observing that the principal eigenfunction inherits the alternating symmetry  of the streamfunction \eref{psi} to satisfy
\beq\elab{phi0a}
 \phi(x+\pi,y)=\phi(x,\pi -y).
 \eeq

Figure \fref{g} shows an instance of $g(c)$ (here for $\Pe=250$) obtained numerically by computing $f(q)$ on a grid in $q$, then Legendre transforming  (the numerical method is described in section  \sref{num}).
Clearly, $g$ is a non-trivial function of $c$, only well approximated by a quadratic function -- corresponding to a diffusive approximation -- in the immediate vicinity of $c=0$. 
We derive below large-$\Pe$ expressions for $g$ that cover the entire range of $c$ and, correspondingly, expressions for the speed $c$ that cover the entire range of $\Da$. 
This requires to analyse three distinguished regimes defined by distinct distinguished scalings of $q$, $c$ and $\Da$.

 \begin{figure*}
	 \begin{center}
 \begin{minipage}{1\linewidth}
	 \hspace{0.11\linewidth}
 \begin{minipage}{0.5\linewidth}
 \begin{overpic}[width=1\linewidth]{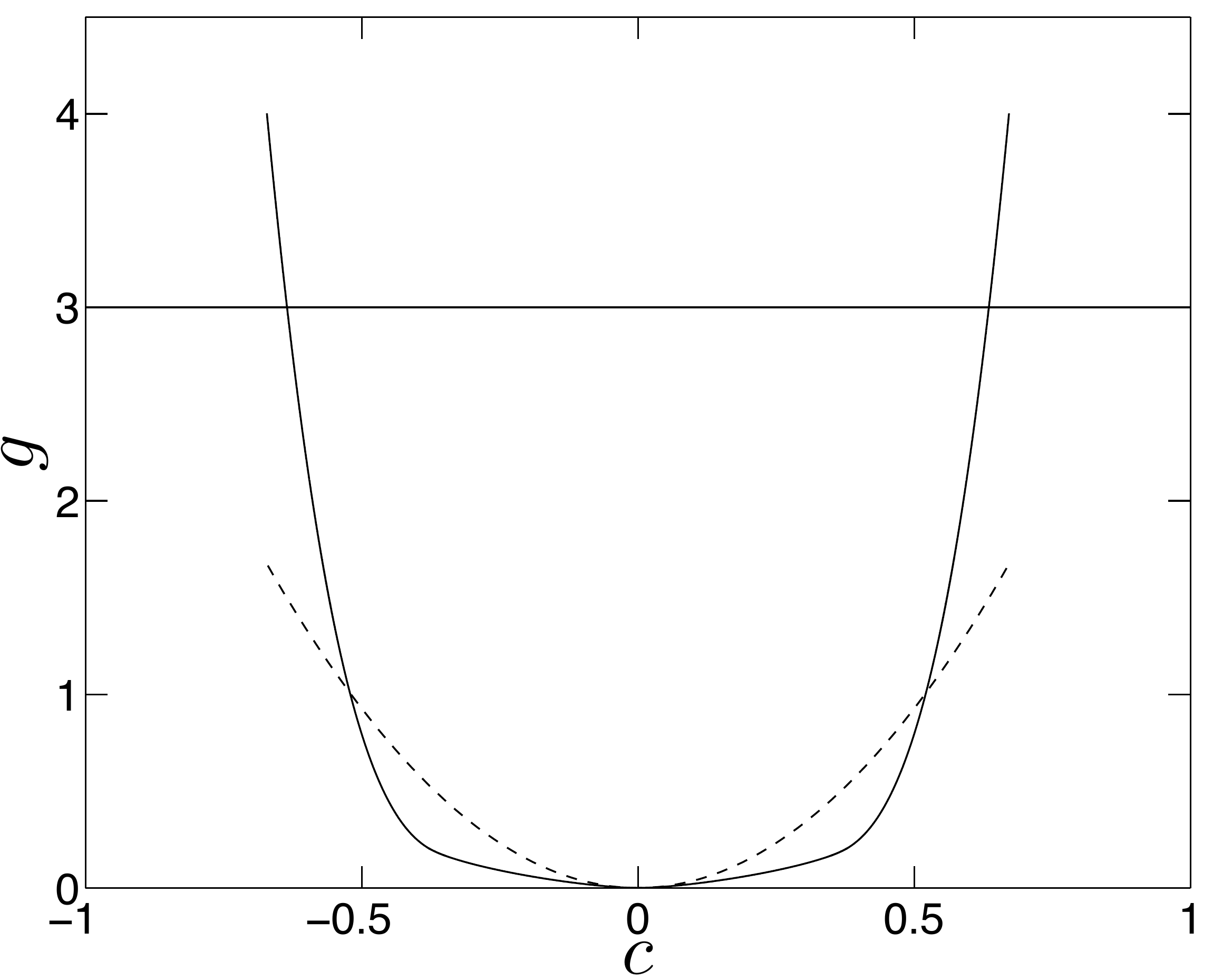}
 \put(85,58){\textcolor{black}{$\Da$}}
 \end{overpic}	
 \end{minipage}
 \begin{minipage}{0.28\linewidth}
	\vspace{2.255cm}
 \centerline{\includegraphics[width=1\linewidth]{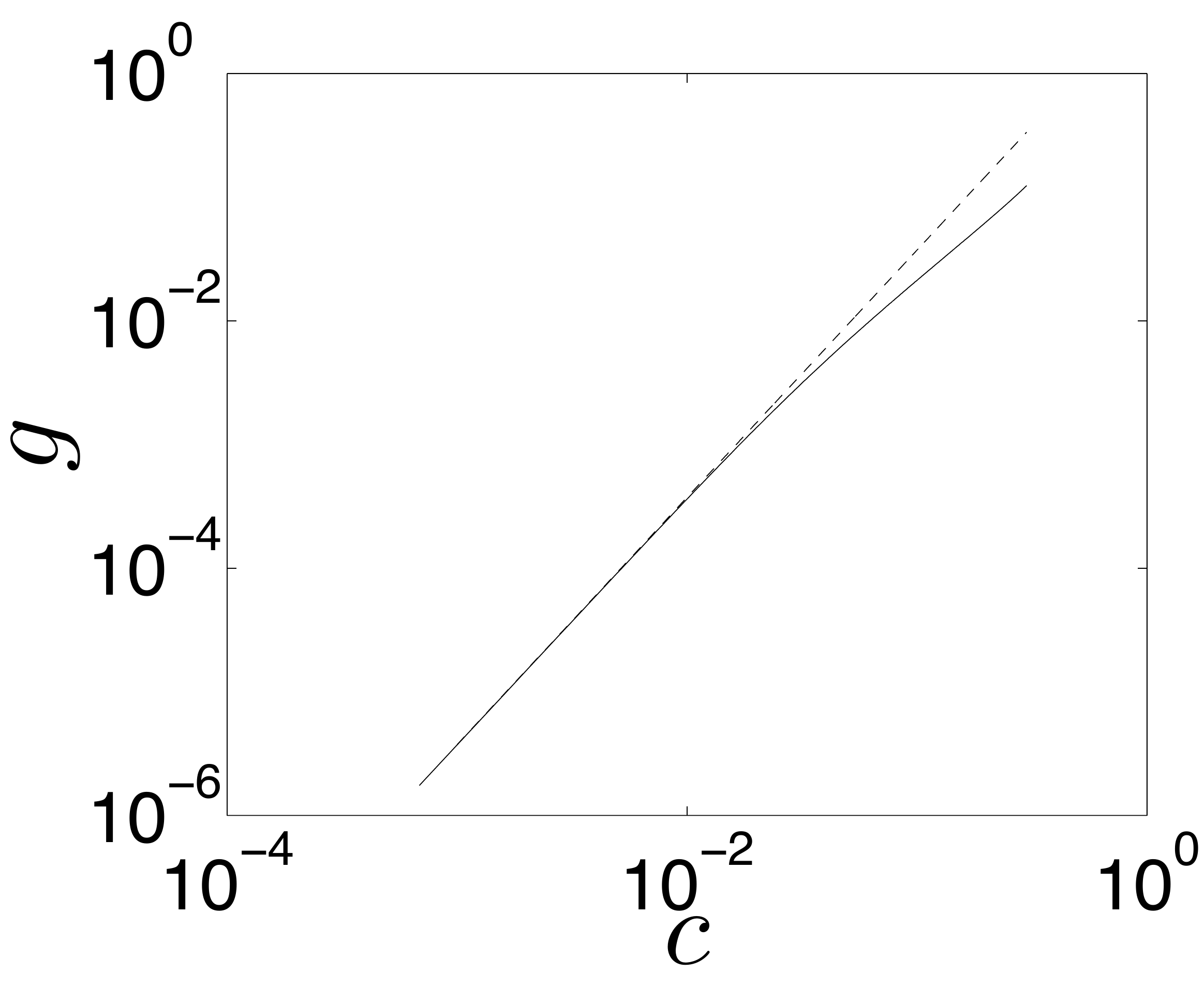}}
 \end{minipage}
  \end{minipage}
   \end{center}
    \begin{center}
   \begin{minipage}{1\linewidth}
	    \hspace{0.11\linewidth}
  \begin{minipage}{0.5\linewidth}
 \centerline{(a)}
 \end{minipage}
 \begin{minipage}{0.28\linewidth}
	  \centerline{(b)}
 \end{minipage}
 \end{minipage}
 \end{center}
 \caption{(Color online). (a) Rate function $g$   calculated numerically for  $\Pe=250$ using \eref{eval2} and \eref{leg}
 	(solid, black line). The diffusive approximation obtained from \eref{eff} is also shown (dashed line).
 	This approximation is equivalent to that derived using
 	homogenization theory
 	and is clearly only valid for $c \ll 1$. (b) Same as (a) but focussing on small values of $c>0$. 
 }
 \flab{g}
 \end{figure*}

\section{Regime I: $\Da=\bm{O(}\Pe^{-1}\bm{)}$}\slab{RegimeI}

The first regime encompasses  the limit of $\Da\to 0$ which is usually tackled using homogenization theory (see e.g. \cite{Bensoussan_etal1978,MajdaKramer1999,PavliotisStuart2007}). 
Homogenization approximates the advection--diffusion equation for a passive scalar by a diffusion equation, in which an effective diffusivity  $\kappa_{\text{eff}}$ replaces molecular diffusivity. This approximation assumes that  $x=O(t^{1/2})$ for $t \gg 1$ and implies that 
\beq\elab{eff}
 g(c)\sim\frac{1}{4}\,\Pe\,\kappa_{\text{eff}}^{-1}c^2\quad\text{and}\quad f(q)\sim \Pe^{-1}\kappa_{\text{eff}}\,q^2,
\eeq 
for $c\ll 1$ and $q\ll 1$ (see \eref{c}). 
For $\Pe \gg 1$, the effective diffusivity for the cellular flow
\cite{Childress1979,Shraiman1987,Rosenbluth_etal1987,Soward1987} is 
\beq\elab{effective}
 \kappa_{\text{eff}} \sim 2\nu\, \Pe^{{1}/{2}}, \quad\text{with} \ \  \nu\approx 0.53,
\eeq
and was obtained in closed form in \cite{Soward1987}. Figure \fref{g} confirms the validity of this approximation and demonstrates its limitation to a very small range of $c$.  

Regime I applies to a broader range of $c$. It can be analysed following \cite{HaynesVanneste2014b} by introducing the rescaling
\begin{equation} 
q=\Pe^{-1/4}\hat{q},\quad\text{where} \ \  \hat{q}=O(1), \elab{tran1}
\end{equation}
as suggested by the form \eref{eff} of $f(q)$ as $q \to  0$.
The eigenvalue and eigenfunction are then expanded according to
\begin{subequations} \elab{exp}
\begin{align}
f(q) &=\Pe^{-1}f_0(\hat{q})+O(\Pe^{-{5}/{4}}),  \elab{expa}\\
\phi& =\phi_{0}+ \Pe^{-{1}/{4}} \phi_1 + \Pe^{-1/2} \phi_2 + \Pe^{-3/4} \phi_3 + \Pe^{-1} \phi_4 +O(\Pe^{-{5}/{4}}). \elab{phiexp}
\end{align}
\end{subequations} 
It is convenient to use the value of the streamfunction $\psi$ and the arclength $s$ along streamlines as coordinates alternative to $(x,y)$. Substituting \eref{exp} into 
\eref{eval2} and using that $\partial_s\bx=\nor{\bu}^{-1}\bu$, we obtain the sequence of problems  
 \begin{subequations}\elab{interior_I}
 \begin{align}
 	&\partial_s \, \phi_0=0, \elab{phi00}\\
 	&\partial_s \, \phi_i=\hat{q} \,\partial_s x\,\phi_{i-1},  \ \ k=1,2,3,\elab{phi0kapp}\\
 	&\nor{\bu}^{-1}\Delta\phi_{0}-\partial_s\phi_4+\hat{q}\,\partial_s x\,\phi_3=\nor{\bu}^{-1}f_0\phi_{0}. \elab{phi04}
    \end{align}
 \end{subequations}
It follows that  $\phi_{0}=\phi_0(\psi)$ is constant along streamlines and  automatically satisfies condition 
\eref{phi0a}. The functions $\phi_i$ for $i=1,2,3$ are polynomials in $x(\psi,s)$ of degree $i$ with $\psi$-dependent coefficients. They do not satisfy \eref{phi0b} and \eref{phi0a}, but these are restored through boundary layers at $x=0,\, \pi$ and $y=0,\, \pi$ which we treat below. Integrating \eref{phi04} around a streamline leads to the 
solvability condition 
\begin{subequations} \elab{int}
\beq
	\dt{}{\psi} \left (a(\psi)\dt{\phi_{0}}{\psi}\right)=f_0\,\phi_{0}\,b(\psi). 
	\eeq
In this equation, derived using that $\frac{\dd}{\dd\psi}\oint_\psi \nor{\nabla\psi}\, \dd s=\oint_\psi\Delta\psi\nor{\nabla\psi}^{-1} \dd s$ \cite{RhinesYoung1983}, $a(\psi)$ and $b(\psi)$ are the circulation and period of orbiting motion along the streamline $\psi$; they are given explicitly by
\beq
a(\psi) 
=8(\text{E}'(\psi)-\psi^2\text{K}'(\psi))  
\quad
\text{and}
\quad
b(\psi)=4\text{K}'(\psi), 
\eeq
\end{subequations}
where $\text{K}'$ and $\text{E}'$ are the complete elliptic integrals of the first and second kind  \cite{NIST:DLMF}. Note that \eref{int} is analogous to an effective diffusion equation obtained by averaging \cite{RhinesYoung1983,FreidlinWentzell1984,Pauls2006}.  

 Equation \eref{int} can  be integrated from the centres $\psi = \mp 1$ of the half-cells outwards. Here we need to distinguish two types of half cells: the `$+$' half-cells, rotating counterclockwise with $\psi=-1$ at their centre and exemplified by $(x,y) \in [0,\pi] \times [0,\pi]$; and the `$-$' half-cells, rotating clockwise with $\psi=1$ at their centre and exemplified by $(x,y) \in [\pi,2\pi] \times [0,\pi]$. Using systematically the upper (lower) signs for `$+$' (`$-$') half-cells, we write the boundary conditions at the centre as
\beq\elab{bc_I}
 \phi_{0}=1\quad\text{and}\quad\phi_{0}^{-1}\dt{\phi_{0}}{\psi}=\pm \frac{f_0}{2}\quad\text{at $\psi=\mp 1$.}
\eeq
The first condition fixes an arbitrary normalisation for $\phi_{0}$ (because \eref{int} is linear); the second ensures that $\phi_{0}$ remains bounded as $\psi \to \mp 1$ (see \cite{HaynesVanneste2014b} for details). 
The solution for $\psi \to 0$  determines the Dirichlet-to-Neuman map $\mathscr{F}(f_0)$, defined as
\beq\elab{deriv1}
\lim_{\psi\rightarrow 0^{\mp}}\phi_{0}^{-1}\dt{\phi_{0}}{\psi} =\pm \mathscr{F}(f_0). 
\eeq
Since $\mathscr{F}(f_0) \not= 0$, $\phi_0$ has a discontinuous first derivative across the separatrix $\psi=0$. This is resolved by a boundary layer which we examine next.

Inside the boundary layer, we use the rescaled variables introduced by \cite{Childress1979}, 
\beq\elab{var}
\zeta=\mp \Pe^{1/2}\psi\quad\text{and}\quad\sigma=\int_0^s\nor{\nabla\psi} \, \dd s,
  \eeq
where $\zeta$ is a rescaled streamfunction whose sign is chosen 
so that $\zeta>0$ in the interior of the $\pm$ half-cells. Note that $0 \le \sigma < 8$ and that the cell corners correspond to $\sigma=0,\, 2,\, 4,\, 6$.
We denote by  $\Phi(\zeta,\sigma)$ the eigenfunction in the boundary layer, and expand this in powers of $\Pe^{-1/4}$ as in \eref{phiexp}. To leading order $\Phi_0$ is a constant, matching the interior solution: $\Phi_0 = \phi_0(0)$. The higher-order terms $\Phi_i$, $i=1,2$ satisfy forced heat equations, with $\sigma$ the time-like variable. Solving these in exactly the manner used in the computation of $\kappa_\mathrm{eff}$  \cite{Childress1979,Shraiman1987,Rosenbluth_etal1987,Soward1987} 
leads to the boundary-layer counterpart of \eref{deriv1}, namely
\beq\elab{deriv2}
\lim_{\zeta\to\infty}\dpar{\Phi_{1}}{\zeta}=0  
\quad\text{and}\quad
\lim_{\zeta\to\infty} {\Phi_0^{-1}} \dpar{\Phi_{2}}{\zeta}=-\frac{\pi^2\nu}{4}\hat{q}^2,
\eeq
A derivation is sketched in Appendix \sref{RegimeIA}.
The matching of the derivative of $\phi$ is ensured to leading order provided that
\beq
\lim_{\psi\rightarrow 0^{\pm}}\phi_{0}^{-1}\dt{\phi_{0}}{\psi} = \mp \lim_{\zeta\to\infty} \Phi_0^{-1} \dpar{\Phi_{2}}{\zeta}.
\eeq
Equating the right-hand sides of \eref{deriv1} and \eref{deriv2}  then yields 
\beq\elab{F}
f_0(\hat{q})=\mathscr{F}^{-1}\left(\frac{\pi^2\nu}{4}\hat{q}^2\right), 
 \eeq
where $\mathscr{F}^{-1}$ denotes the inverse of $\mathscr{F}$. Recalling the scaling $f(q)\sim \Pe^{-1} f_0(\hat{q})$, the above expression gives the asymptotic form of $f(q)$ in Regime I.  
Note that this expression is the same as that obtained previously in  \cite{HaynesVanneste2014b}  for an unbounded domain: the difference in boundary conditions arising from the presence of walls at $y=0,\, \pi$ turns out to be unimportant in this regime.

The front speed is now determined using   \eref{speed}. 
From \eref{exp} and \eref{F}, we deduce that  
\beq
g(c)=\Pe^{-1}\mathscr{G}_{1}(\Pe^{3/4}c)+O(\Pe^{-5/4}) 
\elab{expg1}
\eeq
 where $\mathscr{G}_{1}$ is the Legendre transform of $\mathscr{F}^{-1}$. 
Solving \eref{speed} then gives
\beq\elab{speed_regI}
  c\sim\Pe^{-3/4}\mathscr{C}_{1}(\gamma) \quad\text{for} \ \  \gamma=\Da\,\Pe=O(1), 
 \eeq
where $\mathscr{C}_{1} = \mathscr{G}_{1}^{-1}$. Note that, although this expression is derived assuming formally that $\gamma=O(1)$, it will become clear from our analysis of Regime II below that it applies for the larger range 
$\gamma\ll \Pe(\log\Pe)^{-1}$.

\begin{figure}
			\begin{center}
       	\begin{overpic}[width=0.4\linewidth]{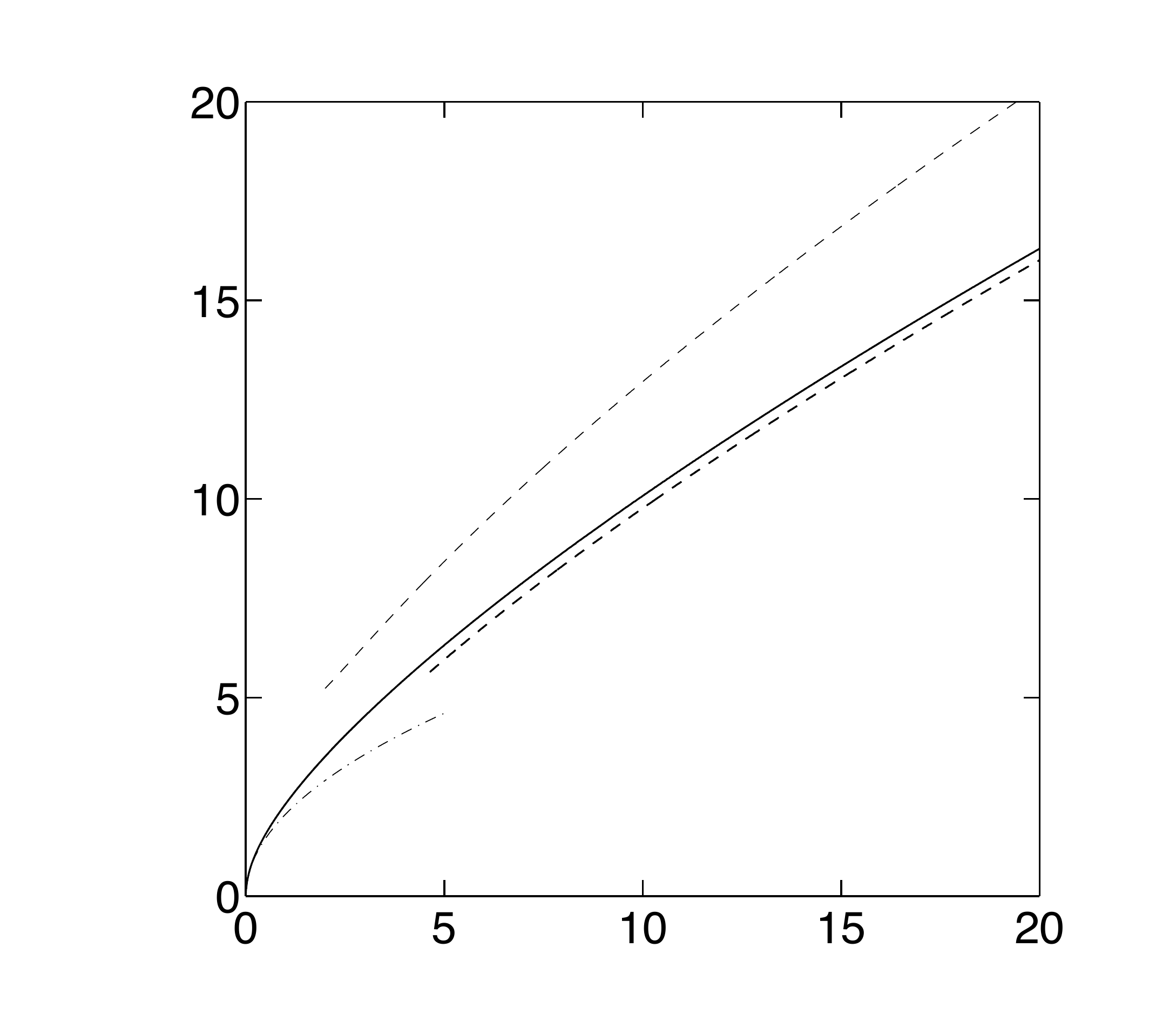}
			\put(-7,42){\rotatebox{90}{\large\textcolor{black}{$\mathscr{C}_1(\gamma)$}}}
			\put(50,-5){\rotatebox{0}{\large\textcolor{black}{$\gamma$}}}
			\end{overpic}	
		\end{center}
\caption{
	Large-$\Pe$ prediction \eref{speed_regI} for the speed $c$ scaled by $\Pe^{3/4}$ as a function of $\gamma=\Pe\Da$.
	The function $\mathscr{C}_{1}(\gamma)$ in \eref{speed_regI} is derived by inverting the Legendre transform of $f_0$ (solid black line) which is obtained from \eref{F} by solving an ODE (\eref{int} with boundary conditions \eref{bc_I}). 
	The small-$\gamma$ approximation \eref{C1_app}  (dot-dashed
	line) is also shown, along with the large-$\gamma$ approximation obtained  (i) by  solving \eref{large_f0}  numerically (lower dashed line)
	and (ii) using the cruder approximation \eref{C1_app2} (upper dashed line). 
}
\flab{C1}
\end{figure}

 Eq.\ \eref{speed_regI} shows that for a fixed value of $\gamma$, 
and thus for constant 
front thickness (since in the absence of advection, this thickness is $(\kappa \tau)^{1/2} =\ell \gamma^{-1/2}$),  $c\propto\Pe^{-3/4}$, which explains the power law that was previously 
conjectured in \cite{Audoly_etal2000} and observed in the numerical work of \cite{Vladimirova_etal2003}. 
It  is also consistent with the rigorous upper and lower bounds scaling as $\Pe^{-3/4}$ obtained in \cite{NovikovRyzhik2007} under the assumption that $\gamma=O(1)$.
It is straightforward to  determine $\mathscr{C}_{1}$ numerically and thus obtain an approximation for $c$. 
We first calculate $\mathscr{F}(f_0)$ for gridded values of $f_0$
using  standard second-order finite differences to discretize \eref{int}  with   boundary conditions \eref{bc_I}. 
Inverting gives $\mathscr{F}^{-1}$ then, by Legendre transforming, $\mathscr{G}_{1}$. Another inversion finally yields  
$\mathscr{C}_{1}$. The result is shown in Figure \fref{C1}. This demonstrates that $\mathscr{C}_{1}$ is a non-trivial function of its argument, implying that a power-law approximation is  only  valid locally.

\subsection*{Asymptotic limits} We now derive  two approximations for $\mathscr{C}_1$  that result in two asymptotic subregimes Ia and Ib of Table \ref{tbl2}. 
Both approximations are based on the asymptotic form of $\mathscr{F}(f_0)$ that 
\cite{HaynesVanneste2014b} derived for small and large values of $f_0 \sim \Pe f$. 
The first approximation uses that
\beq
\mathscr{F}(f_0)=\frac{\pi^2}{8}f_0+O(f_0^2) \quad \textrm{as} \ \ f_0\to 0. \elab{small_f0}
\eeq
Introducing into \eref{F} recovers the quadratic approximation \eref{eff} for $f_0(\hat{q})$. 
We  employ \eref{expg1} and \eref{speed_regI} to deduce that 
\beq
\mathscr{C}_{1}(\gamma)\sim (8\nu \gamma)^{1/2}\quad \textrm{as} \ \  \gamma\to 0. \elab{C1_app}
\eeq
Figure \fref{C1} confirms the validity of this approximation. Eq.\ \eref{speed_regI} then gives the front speed as 
\beq\elab{speed_regIa}
c\sim  
(8\nu)^{1/2}\Da^{1/2}\Pe^{-1/4}\quad\textrm{for} \ \ \Da\ll\Pe^{-1}.
 \eeq
The validity of this approximation  was previously established in \cite{RyzhikZlatos2007,Zlatos2010} where it was shown that in the limit of $\Da\to 0$,  the front speed  is calculated from the quadratic approximation \eref{eff}.

The second approximation uses that
\beq\elab{large_f0}  \mathscr{F}(f_0)=\frac{\sqrt{2}\lambda}{4}\left(1+\frac{\mu}{\log\lambda}\right)+O((\log\lambda)^{-1}) \quad\text{as} \ \ f_0\to \infty,
  \eeq
where $\lambda$ is the solution of  $\lambda^2=4f_0\log\lambda$ and $\mu\approx 0.81$.
 Figure \fref{C1} shows that the corresponding approximation for $\mathscr{C}_{1}$ --  obtained by numerical evaluation of 
 \eref{large_f0}, inversion and Legendre transform  -- is very accurate when its argument is sufficiently large. 
We emphasise that this approximation, although it requires numerical computations, is much simpler than \eref{F}  in that it requires only the solution of algebraic equations instead of the solution of a differential equation.
A closed-form expression is deduced 
by solving
the transcendental equation defining $\lambda$  asymptotically to obtain the leading-order approximation
\beq
  \mathscr{F}(f_0)\sim \frac{1}{2}(f_0\log f_0)^{1/2} \quad\textrm{as} \ \ f_0\to \infty, 
 \eeq
noting that the second term in \eref{large_f0} is subdominant. This approximation is crude because it ignores terms that are $O((\log f_0)^{-1})$ relative to the term retained. 
It is nonetheless useful because it leads to an explicit expression for the speed:
using \eref{F} gives that
$f_0\log f_0\sim \pi^4\nu^2\hat{q}^4/4$ and hence, to leading order, 
that $f_0\sim\pi^4\nu^2\hat{q}^4/(16\log\hat{q})$ as $\hat{q}\to \infty$.
Ultimately,  using $f(q)\sim \Pe^{-1} f_0(\hat{q})$, this gives
\beq\elab{C1_app2}
\mathscr{C}_{1}(\gamma)
 \sim \pi\nu^{1/2}\left({4/3}\right)^{3/4}\gamma^{3/4}(\log \gamma)^{-1/4}
\quad\textrm{as} \ \ \gamma\to \infty.
   \eeq
This expression captures the asymptotic behaviour of $\mathscr{C}_{1}(\gamma)$ but, 
as Figure \fref{C1} shows,  the logarithmic corrections that it neglects are substantially large  for finite $\gamma$.  
Using \eref{speed_regI}, we deduce the  approximation
\beq\elab{speed_regIb}
c\sim
\pi\nu^{1/2}
\left(
{4}/{3}
\right)^{3/4}
\Da^{3/4}(\log\Pe)^{-1/4}
\quad\textrm{for} \ \ \Pe^{-1}\ll\Da\ll(\log\Pe)^{-1},
\eeq
where the upper bound corresponds to the distinguished limit of $\Da$ associated with Regime II. Note that we have dropped a term in $\log\Da$ using that $\Pe\gg\Da$ and $\Pe\gg\Da^{-1}$.
Expression \eref{speed_regIb} will be used below to verify the matching between regimes I and II.

\section{Regime II: $\Da\bm{=O(1/\log\Pe)}$}\slab{RegimeII}

This second regime applies to values of $\Da$ larger than in Regime I which it continues smoothly.
The analysis, which again involves boundary layers, is similar to that carried out in \cite{HaynesVanneste2014b} for the non-reacting problem. 
There are however major differences stemming from the bounded domain that we consider; we therefore describe the analysis in some detail.   

Motivated by the observation that $f=O(q^4)$ when  $q\gg \Pe^{-1/4}$ (up to logarithmic terms, see 
the discussion preceding \eref{C1_app2}),
we assume that $q=O(1)$ and expand the eigenvalue and eigenfunction as
\beq\elab{exp2}
f(q)=f_0(q)+O(\Pe^{-1/4})
\quad\text{and}\quad
\phi=\phi_{0}+O(\Pe^{-1/2}). 
\eeq
Introducing  \eref{exp2} into the eigenvalue equation \eref{eval2}, we find that the interior solution vanishes at leading order: $\phi_{0}=0$. Thus the solution is entirely determined by the behaviour in the boundary layer around the separatrix, as the numerical simulations hint (see  Figure \fref{fronts}(b)).

The boundary layer has a  thickness $O(\Pe^{-1/2})$, as in Regime I; inside, the leading-order solution satisfies
\beq\elab{eval_regimeII}
\partial_{\zeta\zeta}^2\Phi_{0}-\partial_\sigma\Phi_{0}=\frac{f_0-u_1 q}{\nor{\bu}^2}\Phi_{0},
 \eeq
where $\Phi_0$ is expressed in terms of the rescaled variables \eref{var}. 
This can be turned into a heat equation along each segment of the boundary layer using the piecewise transformation 
\begin{subequations} \elab{heat2}
\beq
\widehat{\Phi}=\exp\left(
-q x-f_0\,H(\sigma)
\right)
\Phi_{0},\quad
\text{where} \  \
H(\sigma)=-\int_{2 \lfloor \sigma/2 \rfloor}^\sigma \nor{\bu}^{-2}\, \dd \sigma',
\elab{amp_hat}
\eeq
which reduces \eref{eval_regimeII} to 
\beq
\partial_\sigma\widehat{\Phi} = \partial_{\zeta\zeta}^2\widehat{\Phi}.
\elab{heat}
\eeq
 \end{subequations}
This transformation breaks down near the cell corners where $\nor{\bu}$ vanishes. There, different rescaled variables, namely $(X,Y)=\Pe^{1/4}(x,y)$, are required to solve \eref{eval2}. The solution that is obtained to leading order, namely $\Phi_0  = X^{-f_0} \tilde \Phi(XY)$ for some function $\tilde{\Phi}$, can be matched with the solution of \eref{heat2} upstream and downstream of the corner. This leads to  jump conditions at each corner reading
\beq\elab{jump}
\lim_{\sigma\rightarrow k^+} \widehat{\Phi}(\zeta,\sigma)
=(16\Pe)^{-f_0/2}\zeta^{f_0}
\lim_{\sigma\rightarrow k^-} \widehat{\Phi}(\zeta,\sigma),\quad
\textrm{for} \ \ k=0, \, 2, \, 4, \, 6,
 \eeq
(see \cite{HaynesVanneste2014b} for details).
Combining these jump conditions with (i) the relation between $\hat \Phi$ downstream of each corner and $\hat \Phi$ upstream of the next corner that follows from \eref{heat}, and (ii) the symmetry \eref{phi0a} and boundary conditions \eref{phi0b} results in the eigenvalue problem 
\beq
(16\Pe)^{f_0/2}\widehat{\bm\Phi}(\zeta)=(\mathbfcal{K}\widehat{\bm\Phi})(\zeta)
\elab{sys} \\
\eeq
(see Appendix \sref{appII}).
Here $\widehat{\bf\Phi}$ is a vector grouping the four solutions downstream of each corner, that is, $\hat \Phi(\zeta,\sigma)$ for $\sigma=0^+, \, 2^+,\, 4^+,\, 6^+$,  and  $\mathbfcal{K}$ is a $4\times 4$ matrix operator that depends explicitly on $q$ and $f_0$. Its entries are
linear combinations of the linear integral operators  $\H^\pm$ defined by
\beq
(\H^\pm \Phi)(\zeta)=\frac{1}{\sqrt{8\pi}}\int_0^\infty 
\ee^{-(\zeta\mp\zeta')^2/8}\Phi(\zeta')\, \dd\zeta',
\elab{heatop}
\eeq
  for an arbitrary function $\Phi$.   

An expression for $f_0$ is now obtained by considering the principal eigenvalue of $\mathbfcal{K}$.  
Let $\lambda$ denote this eigenvalue.  
Introducing into \eref{sys} and solving for $f_0$ gives 
\beq\elab{f0_RegimeII}
 f_0=\frac{2\log\lambda}{\log(16\Pe)},\quad\text{where} \ \ \lambda=\lambda(q,f_0).
   \eeq
Note that even though $\log16$ provides an asymptotically negligible correction to $\log\Pe$, it turns out to be significant for the large-but-finite values of $\Pe$ we consider and is therefore better retained.

Equation \eref{f0_RegimeII} is transcendental. 
It is solved numerically by first discretising  $\mathbfcal{K}$  to find $\lambda(q,f_0)$ as the eigenvalue of a matrix, then solving \eref{f0_RegimeII} iteratively, using the straightforward scheme
\beq\elab{f0_RegimeIIappcorr}
 f_{0}^{(n)} =\frac{2\log\lambda\left(q,f_{0}^{(n-1)}\right)}{\log(16\Pe)}, \quad n=1,2,\cdots,
\eeq
taking  $f_{0}^{(0)}=0$ as  initial guess.
 This guess is reasonable when $q\ll 1$
 in which case $f_0\ll 1$. As the value of $q$ increases, the sequence of corrections generated by \eref{f0_RegimeIIappcorr}
 become increasingly important, and increasingly larger values of $\Pe$ are needed
 for the leading-order approximation $f_{0}^{(1)}$ to be accurate.
 
The front speed can be derived from the solution $f_0=f_0(q,\Pe)$ to \eref{f0_RegimeII} by Legendre transforming with respect to $q$ to obtain $g(c)$, then solving $g(c)=\Da$. This leads to $c$ as a transcendental function of $\Da$ and $\log(16 \Pe)$ that can approximated numerically, starting with the  estimate for $f_0$ 
obtained by iterating \eref{f0_RegimeIIappcorr}. 
This approach does not make explicit the scaling relation that characterises Regime II, however. To obtain this, we approximate $\lambda(q,f_0)$ in \eref{f0_RegimeII} by $\lambda(q,0)$, leading to $f_0 \sim f_0^{(1)} = 2 \log \lambda(q,0)/\log(16\Pe)$, and hence to
\beq
g(c)
\sim \frac{\mathscr{G}_{2}(\log(16\Pe)c)}{\log(16\Pe)},  
\elab{expg2}
  \eeq
where  $\mathscr{G}_{2}$ denotes the Legendre transforms of $2\log\lambda(q,0)$ with respect to $q$. The front speed asymptotics 
 \beq \elab{speed_regII}
 c\sim\frac{\mathscr{C}_{2}(\gamma)}{\log(16\Pe)}  
 \quad\text{for}\ \ \gamma=\log(16\Pe)\,\Da,
  \eeq
where   $\mathscr{C}_{2}\equiv\mathscr{G}_{2}^{-1}$, follows. We emphasise that this approximation is asymptotically consistent for $q=O(1)$ since $f_0 \to 0$ as $\Pe \to \infty$. As we show shortly, its accuracy is poor for finite $\Pe$ and the complete solution to \eref{f0_RegimeII}, which treats $1/\log(16 \Pe)$ as $O(1)$, is preferable.

 Figure \fref{C2} shows the behaviour of $\mathscr{C}_{2}$ obtained numerically for a range of values of  $\gamma=\log(16\Pe)\, \Da$.
 The range is limited because 
 the matrix associated 
  with the discretised version of $\mathbfcal{K}$ (with $f_0=0$) becomes  ill conditioned when $\gamma\gtrsim 1$, leading to numerical inaccuracies in the principal eigenvalue $\lambda(q,0)$.   
The complete solution to \eref{f0_RegimeII} leads to a ($\Pe$-dependent) approximation to $c\log(16\Pe)$ which, in contrast, is well conditioned over a broad range of $\gamma$; this approximation is shown in   Figure \fref{C2} for four values of $\Pe$.
The results indicate that  the logarithmic corrections included in the complete solution are negligible for $\gamma\lesssim 1$, with \eref{speed_regII} providing a good approximation, but  significant for larger $\gamma$ when they are seen to decrease very slowly as $\Pe$ increases. The results are also consistent with the behaviour $c\log(16\Pe) \sim \mathscr{C}_{2}(c) \sim \pi$ for $\gamma \gg 1$ derived below.

\begin{figure}
\begin{center}
	\begin{minipage}{1\linewidth}
		\begin{minipage}{0.33\linewidth} 
	    \begin{center}
			\begin{overpic}[width=0.96\linewidth]{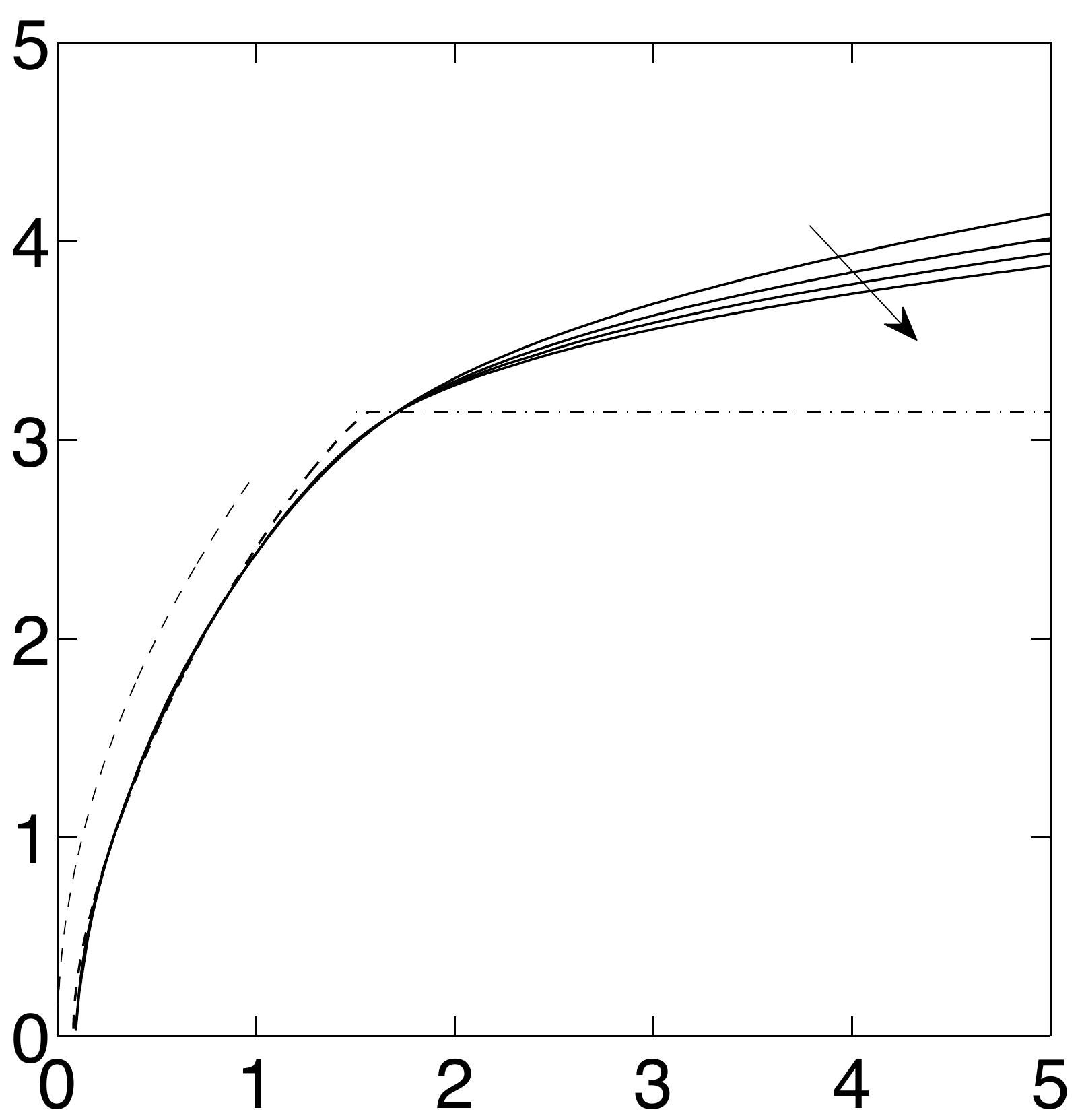}
			\put(-12,42){\rotatebox{90}{\large\textcolor{black}{$\mathscr{C}_2(\gamma)$}}}
			\put(45,-5){\rotatebox{0}{\large\textcolor{black}{$\gamma$}}}
			\end{overpic}	
		\end{center}
		\vspace{0.5cm}	
	     \centerline{(a)}
		 \end{minipage}
		 \hspace{0.2cm}
		 \begin{minipage}{0.66\linewidth} 
	 	    \begin{center}
	 			\begin{overpic}[width=0.98\linewidth]{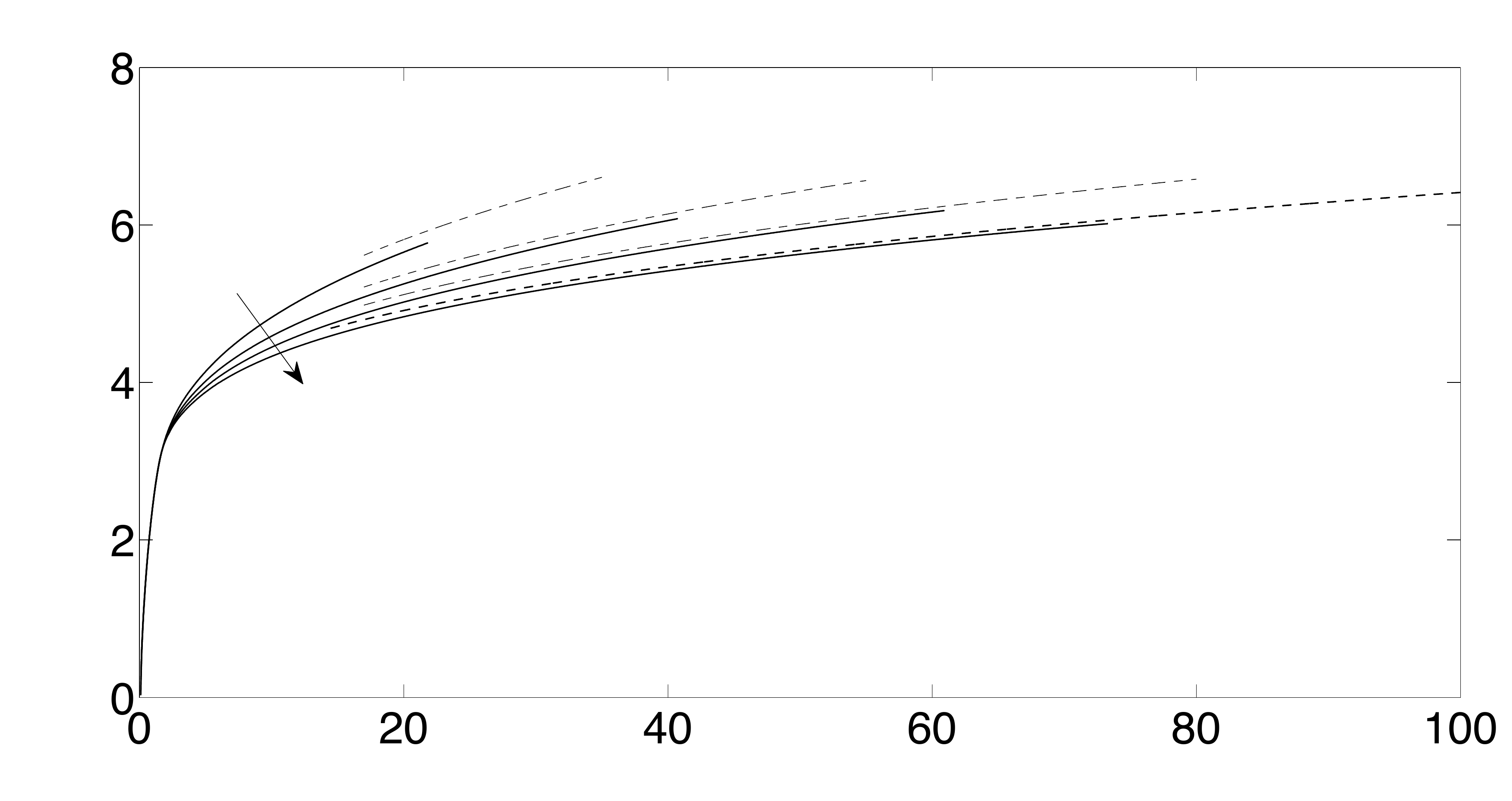}
	 			\put(-5,21){\rotatebox{90}{\large\textcolor{black}{$\mathscr{C}_2(\gamma)$}}}
	 			\put(48,-2){\rotatebox{0}{\large\textcolor{black}{$\gamma$}}}
	 			\end{overpic}	
	 		\end{center}
			\vspace{0.4cm}	
	 	     \centerline{(b)}
	 		 \end{minipage}			 
	\end{minipage}
\end{center}
\caption{
(a) Large-$\Pe$ prediction \eref{speed_regII} for the front speed $c$ scaled by $\log(16 \Pe)$ as a function of $\gamma=\log(16\Pe)\Da=O(1)$ (lower dashed line).
The function $\mathscr{C}_2$ in \eref{speed_regII} is obtained by computing the principal eigenvalue of   \eref{sys}
for $f_0=0$ numerically.
The small-$\gamma$ approximation \eref{C2_app} (upper dashed line) and large-$\gamma$ approximation $\mathscr{C}_2 \sim \pi$ (dashed-dotted line) are also shown.
The four thin solid black lines correspond to higher-order 
corrections to  \eref{speed_regII}  obtained  
for $\Pe=50$, $125$, $250$ and $500$, with the arrow pointing in the direction  of increasing $\Pe$.
(b)  Higher-order corrections  compared to the large-$\gamma$ approximation \eref{speed_regIIbv2} (dashed lines). 
  }
 \flab{C2}
 \end{figure}

\subsection*{Asymptotic limits} There are two asymptotic approximations of the front speed in Regime II, corresponding to $\gamma \ll 1$ and $\gamma \gg 1$ and identified as subregimes IIa and IIb in Table \ref{tbl2}. For the first, we approximate $\mathscr{C}_2$ in \eref{speed_regII}
based on the asymptotic form of $\lambda(q,0)$ for $q\ll 1$  derived in  \cite{HaynesVanneste2014b}. 
For such $q$, the jumps in  \eref{jump} are negligible, and the boundary-layer solution can be expanded in powers of $q$, whence it is found that 
$\lambda(q,0)\sim \exp(2\mu^2)$ where $\mu=\pi^2\nu q^2/4$.
It follows that 
\beq\elab{C2_app}
\mathscr{C}_{2}(\gamma)\sim \pi \nu^{1/2} \left(4/3\right)^{3/4} \gamma^{3/4}\quad\textrm{for} \ \ \gamma\ll 1,
\eeq
and, using \eref{speed_regII}, that the front speed is that reported in Table \ref{tbl2}. This Regime IIa asymptotic expression coincides with that found in Regime Ib as \eref{speed_regIb}, thus confirming the matching between Regimes I and II. 

The second asymptotic approximation corresponds to $\gamma\gg 1$, hence $q\gg 1$. In this limit, the eigenvalue $\lambda(q,f_0)$ of $\mathbfcal{K}$ can be derived from a scalar eigenvalue problem which we derive and solve asymptotically in Appendix \sref{f0_asym_large}. From this solution, we deduce the asymptotics 
\eref{large_f0II} for $f_0$. Taking the Legendre transform gives
$g(c)\sim 8 \Pe \, c \, \ee^{-\pi/c}/\pi$,
which we invert to obtain the front speed in Regime IIb as 
\beq\elab{speed_regIIbv2}
c\sim \frac{\pi}{\text{W}_\mathrm{p}(8\Pe\Da^{-1})} \quad\textrm{for}  \ \ \frac{1}{\log\Pe}\ll\Da\ll\Pe,
\eeq
where $W_\mathrm{p}$ denotes the principal real branch of the Lambert W-function (solution of  $W(z)\, \ee^{W(z)}=z$, see \cite{NIST:DLMF})
The upper bound of the range of validity of  \eref{speed_regIIbv2} is determined by comparison with the results in Regime III in the next section. 

Figure \fref{C2} shows that the approximation \eref{speed_regIIbv2}  is very good for $\Pe=50$, $\Pe=125$ and excellent for $\Pe=250$ and $500$.  
We note that, since $\Da\ll\Pe$ and $\Pe \gg \Da^{-1}$, it is consistent to approximate $\text{W}_\text{p}(8\Pe\Da^{-1})$ by $\log \Pe$ \cite{NIST:DLMF} to reduce \eref{speed_regIIbv2} to 
\beq\elab{speed_IIb}
c\sim\frac{\pi}{\log\Pe}.
\eeq
This approximation is poor for finite $\Pe$ because of the neglect of logarithmic error terms. 
It is useful in that it shows that both the  small-$\gamma$ and large-$\gamma$  approximations lead to the same scaling \eref{speed_regII} for the front speed, with $\mathscr{C}_2(\gamma) \to \pi$ as $\gamma \to \infty$.

We note that an expression qualitatively similar to \eref{speed_IIb}  was obtained in \cite{Abel_etal2002,Cencini_etal2003}  using the so called G-equation, a model alternative to (but not derived from) the FKPP model when applied to fast reaction. This expression suggests that the front speed $c$ is independent of $\Da$ for a range of $\Da$; as the more complete approximation \eref{speed_regIIbv2} shows and Figure \fref{C2} confirms, there is in fact a slow growth of $c$ with $\Da$. This growth is actually logarithmic, as can be made explicit by improving the approximation of \eref{speed_regIIbv2} to include the first-order correction to \eref{speed_IIb} and obtain
\beq
c\sim \frac{\pi}{\log\Pe}   + \frac{\pi \log \Da}{\log^2\Pe}.
\eeq

\section{Regime III: $\bm{\Da=O(\Pe)}$}\slab{RegimeIII}
This final regime corresponds to a fast reaction and may be referred to as a geometric-optics regime. 
Our analysis of Regime IIb (and specifically,  \eref{large_f0II}) suggests that Regime III emerges for $q=O(\Pe)$. 
We therefore introduce the rescaling
\beq
q =\Pe\,\hat{q},\quad\textrm{where} \ \ \hat{q}=O(1), \elab{tran3}
\eeq
into the eigenvalue problem \eref{eval2}. We then expand the eigenvalue according to
\begin{subequations}\elab{expansions3}
\beq
f(q)=\Pe f_0(\hat{q}) + O(1)
\eeq
and assume that the eigenfunction takes the WKB form
\beq
\phi=\ee^{-\Pe w} \left( a + O(\Pe^{-1}) \right), 
\elab{exp3}
\eeq
\end{subequations} 
where $w$ and $a$ satisfy the same boundary conditions  as $\phi$. 
Substituting  \eref{expansions3} into  \eref{eval2} leads to
\beq 
H(\nabla w,\bm{x})= f_0,\quad
\text{where $H\equiv\nor{\nabla w}^2 + \bu \cdot \nabla w + 2 \hat q \partial_x w + u_1 \hat q + {\hat q}^2$} 
\elab{whj}
\eeq
can be regarded as a Hamiltonian. 
This nonlinear eigenvalue problem has been obtained for general flows by Freidlin, Evans and Souganidis, and Majda and Souganidis (see \cite{Freidlin1985}, \cite{Evans_etal1989} and  \cite{MajdaSouganidis1994}). It can be interpreted as the cell problem arising in the homogenisation of the  Hamilton--Jacobi equation $\partial_t w + \nor{\nabla w}^2 + \bu \cdot \nabla w = 0$ and  has been shown to have a unique solution $f_0$ for  each value of $\hat{q}$ \cite{Lions_etal}.

Eq.\ \eref{whj} cannot be solved analytically in general, and direct numerical solutions are rather involved (see e.g. \cite{KhouiderBourlioux2002} for the specific case of the cellular flow).
Here we exploit a variational formulation which expresses $f_0$, or rather its Legendre dual, the rate function $g_0$ (such that $g(c) = \Pe \, g_0(c)+O(1)$), in terms of a minimum-action principle. 
We derive this variational formulation by considering a time-dependent version of \eref{whj}, namely the Hamilton--Jacobi equation
\beq
\partial_t w + H(\nabla w,\bm{x})=0,
\elab{hj}
\eeq
noting that we can expect
\beq
f_0(\hat q) = - \lim_{t \to \infty} \frac{w(\bx,t)}{t}
\elab{fff}
\eeq
 for a wide range of initial conditions $w(\bx,0)$. The solution of \eref{hj} can  be written in terms of action-minimising paths $\bvarphi( \cdot) =(\varphi_1(\cdot),\varphi_2(\cdot)) \in \mathbb{R} \times [0,\pi]$, specifically as
\beq
w(\bx,t) = \inf_{\bvarphi(\cdot)} \left\{ \int_0^t  L_w(\dot{\bvarphi}(s),\bvarphi(s)) \, \dd s  \,\bigg\vert\,   \bvarphi(t)=\bx \right\}, \quad \textrm{for} \ \ t > 0,
\elab{w}
 \eeq
assuming that $w(\bx,0)=0$ (e.g., \cite{Evans}). The Lagrangian $L_w$ is derived by taking the Legendre transform of $H$ to find
\beq
L_w(\dot{\bvarphi}(s),\bvarphi(s)) = L(\dot{\bvarphi}(s),\bvarphi(s)) - \hat q \dot \varphi_1(s),
\elab{Lw}
\eeq
where 
\beq
L(\dot{\bvarphi}(s),\bvarphi(s))=\frac{1}{4}\nor{\dot{\bvarphi}(s)-\bu(\bvarphi(s))}^2.
\elab{lagrangian}
 \eeq
 Using \eref{w} and \eref{Lw}, we rewrite \eref{fff} as
 \beq
 f_0(\hat q) = - \lim_{t \to \infty}  \frac{1}{t} \inf_{\bvarphi(\cdot)} \left\{ \hat q  \varphi_1(0) -   \hat q x  
 + \int_{0}^t L(\dot{\bvarphi}(s),\bvarphi(s)) \, \dd s  \,\bigg\vert\, \bvarphi(t)=\bx \right\}.
 \eeq
 Without loss of generality we choose  $x=\phi_1(t)=0$ and leave $y=\phi_2(t)$ undetermined.
 This is possible because changes to  their  values  lead to $O(1)$ changes to the infimum and therefore leave $f_0$ unaffected.
We further make the transformation $\bvarphi(s) \mapsto - \bvarphi(t-s)$. This leaves the Lagrangian \eref{lagrangian} unchanged and enables us to rewrite $f_0$ as
\beq
f_0(\hat q) = \lim_{t \to \infty}  \frac{1}{t} \sup_{\bvarphi(\cdot)} \left\{ \hat q  \varphi_1(t) - \int_{0}^t L(\dot{\bvarphi}(s),\bvarphi(s)) \, \dd s  \,\bigg\vert\, \bvarphi(0)=(0, \cdot) \right\}. 
\eeq
 We now introduce $c =  \varphi_1(t)/t$  to obtain that 
\beq\elab{f0sign}
f_0(\hat q)=\sup_{c} \left( \hat q c -  \lim_{t \to \infty} \frac{1}{t} \inf_{\bvarphi(\cdot)}
\left\{ \int_{0}^t L(\dot{\bvarphi}(s),\bvarphi(s)) \, \dd s  \,\bigg\vert\, \bvarphi(0)=(0,\cdot), \bvarphi(t)= (c t,\cdot) \right\} \right),
\nonumber\eeq
where the dependence on specific values of  
  $\varphi_2(0)$, $\varphi_2(t)$ is dropped. 
Recognizing the Legendre transform,  we obtain the rate function
\beq
g_0(c)= \lim_{t \to  \infty} \frac{1}{t} \inf_{\bvarphi(\cdot)}
\left\{ \int_0^t L(\dot{\bvarphi}(s),\bvarphi(s)) \, \dd s  \,\bigg\vert\, \bvarphi(0)=(0,\cdot), \bvarphi(t)= (c t,\cdot) \right\}.
\elab{abc}
\eeq
This gives $g_0(c)$ in terms of the action-minimising path -- or instanton -- $\bvarphi^*(\cdot)$.

We make three remarks. First, the result \eref{abc} follows directly from an application of the Freidlin--Wentzell (small noise) large-deviation theory
(see \cite{FreidlinWentzell1984},\cite[Ch.~6]{Freidlin1985} and \cite{FreidlinSowers1999})
to the dispersion of passive particles in the flow $\bu$. 
Thus Regime III can be regarded as lying at the intersection between large-$t$ large-deviation theory as used in this paper, and small-noise (large-$\Pe$) large-deviation theory: that their results coincide indicates that the two limits $t \to \infty$ and $\Pe \to \infty$ commute. Second, the asymptotics of the principal eigenvalues of a broad class of second-order elliptic operators can be obtained using a variational approach \cite{Piatnitski1998}; thus, \eref{abc} could be alternatively derived by  application of the relevant results in \cite{Piatnitski1998}. Third, since \eref{whj} is the cell problem for the homogenisation of a Hamilton--Jacobi equation \cite{Evans_etal1989,MajdaSouganidis1994}, \eref{abc} provides a variational route to derive the homogenised Hamiltonian $f_0$.

 Computing the right-hand side of \eref{abc} becomes considerably easier by observing that we may take the minimising path to be periodic, in the sense that 
\beq
\bvarphi(s+\tau)=\bvarphi(s)+(2\pi,0),
  \quad\textrm {where} \ \ \tau={2\pi}/{c}.
\eeq
Using that $\int_0^{n\tau}L(\dot{\bvarphi}(s),\bvarphi(s))\, \dd s= n\int_0^\tau L(\dot{\bvarphi}(s),\bvarphi(s))\, \dd s$  reduces  \eref{abc} to
\beq
g_0(c)= \frac{1}{\tau} \inf_{\bvarphi(\cdot)}
\left\{ \int_0^\tau L(\dot{\bvarphi}(s),\bvarphi(s)) \, \dd s  \,\bigg\vert\, \bvarphi(0)=(0,\cdot), \bvarphi(\tau)= (2 \pi,\cdot) \right\}. 
\eeq
Recalling the scaling $g(c)\sim \Pe \, g_0(c)$ and letting $\sigma=c s$ in the above expression, we finally obtain the rate function as
\beq
g(c)=\Pe\,\mathscr{G}_3(c),  
\eeq
where
\beq
\mathscr{G}_3(c)=\frac{1}{8\pi}\inf_{\bvarphi(\cdot)}
			\left\{
			\int_0^{2\pi}\nor{c{\bvarphi}'(\sigma)-\bu(\bvarphi(\sigma))}^2d\sigma 
			\,\bigg\vert\,\bvarphi(2\pi)=\bvarphi(0)+ (2\pi,0)\right\}.
			\elab{G}		
\eeq
The front speed in Regime III follows as 
\beq\elab{c3}
c\sim\mathscr{C}_3(\gamma)\quad\textrm{for} \ \ \gamma=\Da/\Pe=O(1),
\eeq
where $\mathscr{C}_3\equiv\mathscr{G}_3^{-1}$. The authors derived this result previously using a different approach, directly related to the Freidlin--Wentzell small-noise large deviation, that bypasses the eigenvalue problem \eref{eval2}  \cite{TzellaVanneste2014}. While the present derivation is more involved, it highlights the relation with the eigenvalue problem and hence the connection between the three regimes.

The minimization problem  \eref{G} provides an easy way to compute the instanton and thus the front speed numerically.  
Its solution is  straightforward to obtain using MATLAB's optimization toolbox.	
We first start with a large value of $c$ and use a  standard first-order finite-differences
 to discretize  $\sigma$   in $N=250$ equidistant points. The resulting discrete action is then minimized  using the 
routine {\tt fminunc} that is seeded with the straight line 
 $\bvarphi^\ast(s)=(cs,\pi/2)$ as initial guess.
 We then iterate over a range of values of $c$ using the previously determined  
 path as an initial guess  to find the next minimizer.
Figure 2 in \cite{TzellaVanneste2014} shows characteristic examples of instantons $\bvarphi^\ast(s)$ that are obtained 
 for different values of $c$. 
These are close to a straight line when $c$ is large  
and follow   closely a streamline near the  cell boundaries when $c$ is small.  
Figure \fref{speed_asym} shows the behaviour of $c$ as a function of $\gamma$ deduced from \eref{c3}.

\begin{figure}
\begin{center}
	\begin{overpic}[scale = .29]{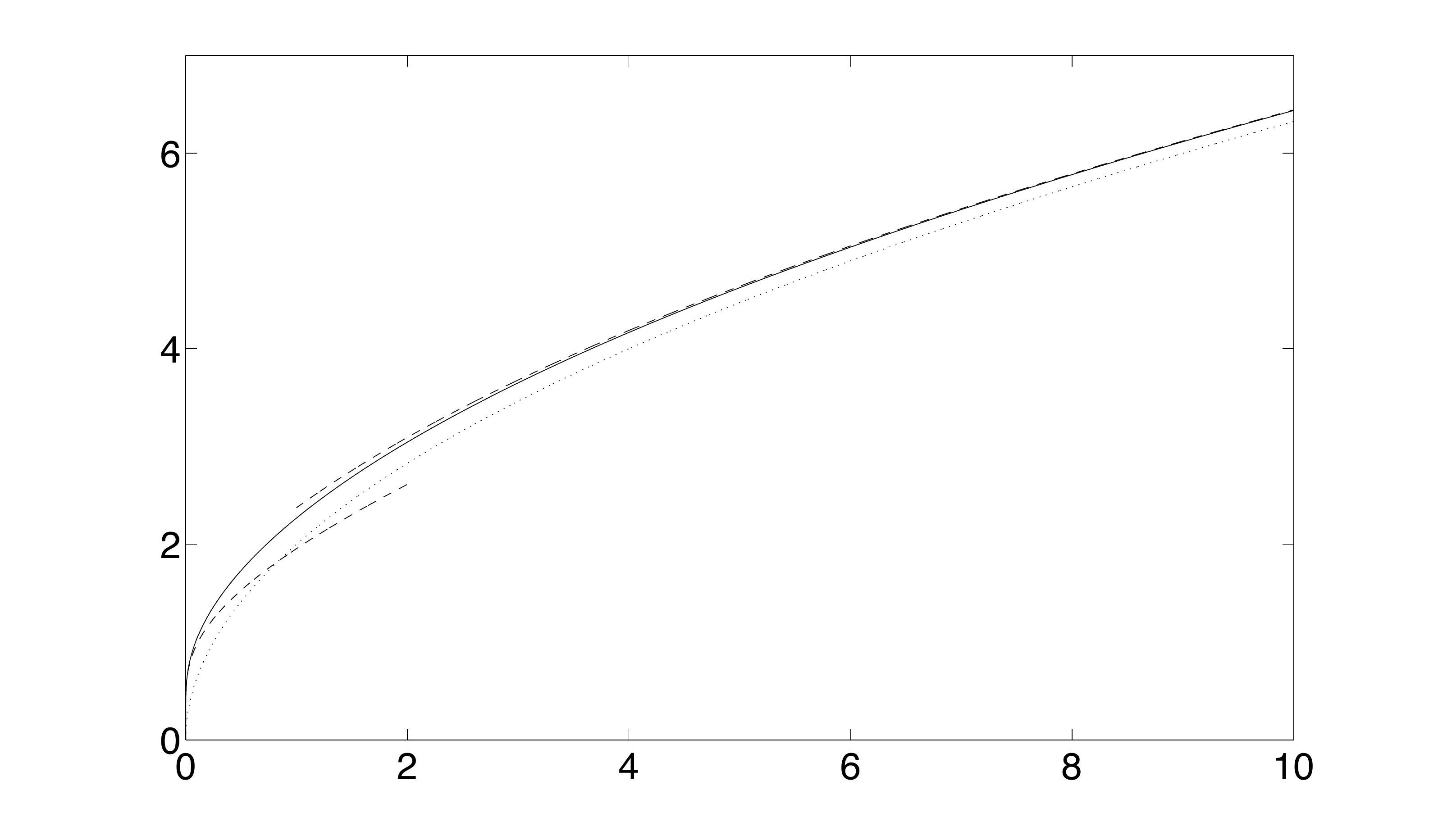}
		\put(50,5){\includegraphics[scale=0.155]{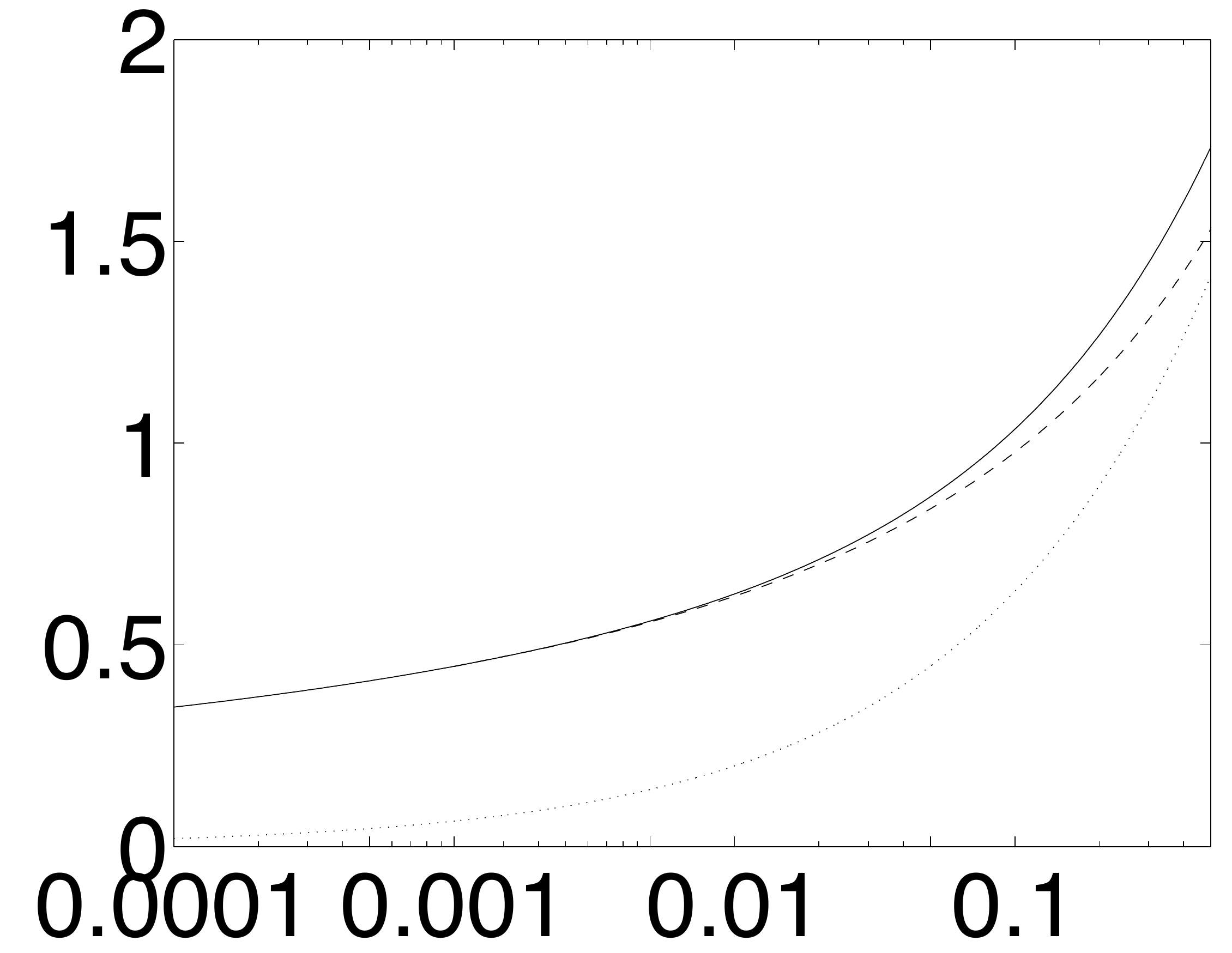}}
		\put(-7,27){\rotatebox{90}{\large\textcolor{black}{$\mathscr{C}_3(\gamma)$}}}
		\put(48.5,-4){\rotatebox{0}{\large\textcolor{black}{$\gamma$}}}
	\end{overpic}	
\end{center}
\caption{
Large-$\Pe$ prediction \eref{c3} for the front speed $c$ 
valid for $\gamma=\Da/\Pe=O(1)$.
$\mathscr{C}_3$ is calculated numerically  by minimizing \eref{G} (solid black line) and compared with 
the small-$c$ asymptotic approximation \eref{speed_regIIbv2} (lower dashed line), the large-$c$ asymptotic approximation \eref{c1app} (upper dashed line), and the bare speed $c_0=2\sqrt{\gamma}$ (dotted line). The inset focuses on smaller values of $\gamma$   (after \cite{TzellaVanneste2014}). 
   }
 \flab{speed_asym}
 \end{figure}

\subsection*{Asymptotic limits}
Closed-form expressions for $c$ are derived in \cite{TzellaVanneste2014} for two asymptotic limits, corresponding to $\gamma \ll1 $ and $\gamma \gg 1$ and referred to as subregimes IIIa and IIIb in Table \ref{tbl2}.
We sketch the derivation here for completeness. 

For  $\gamma \ll 1$ and hence $c \ll 1$, the instanton follows a streamline close to the cell boundaries, departing from it only for $y \approx \pi/2$. The action \eref{G} is minimized when $\bphi^\ast(\sigma)=(x(\sigma),y(\sigma))$ satisfies $cy'\approx -\cos x\sin y$ (so that the instanton and flow speeds differ only in the $x$-direction).  
Exploiting symmetry to consider $0 \le \sigma \le \pi/2$ only, with $x(0)=0$, $y(0)=x(\pi/2)=\pi/2$ and $y'(\pi/2)=0$, we can divide the instanton path  into two segments. In region $1$,  where
     $x\ll 1$, the integrand in \eref{G} is  approximately $(cx'-x\cos y)^2$, leading to the Euler--Lagrange equation $c^2 x''=x$ (since $cy'\approx-\sin y$). 
In   region $2$, $y\ll 1$, $cx'=\sin x\cos y\approx-\sin x$  and $cy'=-\cos x\sin y\approx-y\cos x$.  
Matching between the solutions in their common region of validity $x,y\ll 1$ (the cell corner) gives the approximation
\beq\elab{instantonsmallc}
\bphi^\ast(\sigma)\sim
  \begin{cases}
   \left({C}_1(\sigma),{C}_2(\sigma)\right) & \text{for }\sigma \ll \pi/2 \\
    \left({C}_2(\pi/2-\sigma),{C}_3(\pi/2-\sigma)\right) & \text{for }\sigma \gg c
  \end{cases}
\eeq
where $C_1(\sigma)=4 \exp(-\pi/(2c))\sinh(\sigma/c)$, $C_2(\sigma)=2\tan^{-1}(\exp(-\sigma/c)))$ and 
$C_3(\sigma)=4 \exp(-\pi/(2c))\cosh(\sigma/c)$. 
Expression \eref{instantonsmallc} is in very good agreement with our numerical solution. 
Using \eref{instantonsmallc} gives the integrand in  \eref{G} as
$(cx'-x\cos y)^2\approx 16\exp\left(-\pi/c\right)\cosh^{-2}\left(\sigma/c\right)$, leading to
\beq\elab{G1}
\mathscr{G}_3(c)\sim  4\times(2/\pi)ce^{-\pi/c},\quad\text{where $c\ll 1$}
  \eeq
and the factor $4$ appears because, for  $\sigma \in [0\,\,2\pi]$, the solution \eref{instantonsmallc} repeats 4 times, up to symmetries.
Inverting \eref{G1} yields 
\beq \elab{c1app}
c \sim \frac{\pi}{W_\mathrm{p}(8 \gamma^{-1})} \quad \textrm{for} \ \ \gamma = \Da/\Pe \ll 1,
\eeq
that is, the same expression as \eref{speed_regIIbv2} found as Regime IIb. This verifies the matching between Regimes II and III.

The second asymptotic limit corresponds to $\gamma \gg 1$, hence $c\gg 1$. 
In this case, the instanton path is approximately a straight line, with expansion
\beq
\bvarphi^\ast(\sigma)=(\sigma,y_0)+c^{-1}(x_1(\sigma),y_1(\sigma))+O(c^{-2})
\elab{insc}
\eeq
where $x_1$, $y_1$ are $2 \pi$-periodic functions  satisfying $x_1(0)=y_1(0)=0$. 
Substituting  into \eref{G} and minimising with respect to $y_0$, $x_1(\sigma)$ and $y_1(\sigma)$ gives $x_1(\sigma)=0$, $y_1(\sigma)=-2\sin\sigma\sin y_0$ and $y_0=\pi/2$, leading to
\beq\mathscr{G}_3(c)=c^2/4-3/8 +O(c^{-2}).
\eeq
Using \eref{c3} finally leads to the asymptotics of the speed
\beq\elab{c2app} 
c 
\sim 2\sqrt{\gamma}\left(1+\frac{3}{16\gamma}\right)
\quad \ \ \textrm{for} \ \ \gamma\gg 1. 
\eeq
The leading-order term in \eref{c2app} is  the bare speed $c_0$, unsurprisingly since reaction is so strong in this regime that advection has a small effect on the front evolution.  The second term in the expansion is necessary for a good agreement between  asymptotic and  full results (see Fig. \fref{speed_asym}).

\section{Comparison with numerical results}\slab{num}
We compare our predictions  for the speed $c$ derived in each regime with the corresponding values obtained  from (i) the
numerical evaluation of the principal eigenvalue in \eref{eval2}, and (ii) direct numerical simulations of the FKPP equation \eref{FKPP} with $r(\theta)=\theta(1-\theta)$. 
For (i) we 
use  a standard second-order finite-difference discretization of \eref{eval2}. 
The resulting matrix eigenvalue problem is solved for a range of values of $q$ 
using  MATLAB's routine {\tt eigs}. 
We choose the spatial resolution  $\Delta$ to satisfy   $\pi/\Delta=750$ in both directions.  

For (ii) we discretize   
\eref{FKPP} 
using a fractional-step method with a Godunov splitting 
in which we alternate between independent advection, diffusion and reaction steps. 
The advantage of this method is that it is simple and cheap to combine a high-resolution finite-volume  method for the advection equation  $\partial_t\theta+\bu\cdot\nabla\theta=0$, with an alternating-direction implicit method for the diffusion equation $\partial_t\theta=\Pe^{-1}\Delta\theta$, and an exact solution of the reaction equation $\partial_t\theta=\Da \, r(\theta)$.  
The advection equation  
is solved using a first-order upwind method that includes a minmod limiter to account for second-order corrections  
(see \cite{Leveque} for more details). This is a stable scheme as long as 
the Courant--Friedrichs--Lewy (CFL) condition is satisfied. 
We choose the spatial resolution $\Delta$ to  satisfy
 $\pi/\Delta=400$ when $\Da<1$ and $\pi/\Delta=750$ otherwise. 
This way we ensure that $\Delta/\pi>10^{-1}\text{min}(\delta_1,\delta_2)$  for all values of $\Pe$ and $\Da$  
where $\delta_1=O(\Pe^{-1/2})$ and $\delta_2=O(\Pe^{-1/2}\Da^{-1/2})$   are  the  characteristic thicknesses of the boundary layer and front, respectively. 
The time-step is controlled by the CFL number that we set to be equal to $0.8$.

To make the computational domain finite, we set artificial boundaries at $x=\pm N\pi$, 
with $N=15$ when $\Da<1$ and $N=5$ otherwise, so that boundary effects are negligible. A larger domain is necessary for smaller $\Da$ values because the front width is larger (see, e.g.,  Fig.\ \fref{fronts}(a)).
We impose absorbing boundary conditions using a zero-order extrapolation at each of the four boundaries.  We modify the computational domain to track the  front for a long   time:  
each time the  
 solution  at $x=(N-1)\pi$  
 becomes larger than 
$\varepsilon=10^{-6}$, we eliminate 
the nodes with $-N\pi\leqslant x \leqslant (-N+1)\pi$ to the left of the front 
and add new nodes with 
$N\pi\leqslant x \leqslant (N+1)\pi$ to the right of the front where we set $\theta= 0$.
The front speed is insensitive to the precise value of $\varepsilon$. 
We calculate the speed of the front by considering the left and right endpoints of the front, $x^{-}_\epsilon(t)$ and $x^{+}_\epsilon(t)$, defined as 
\beq\elab{position}
	x^{-}_\epsilon(t)=\text{min}\{x:\theta(x,t)=1-\epsilon\}
	\quad\text{and}\quad
	x^{+}_\epsilon(t)=\text{max}\{x:\theta(x,t)=\epsilon\},  
\eeq
which we determine using a third-order polynomial interpolation. 
We calculate the large-scale speed of the front from a linear fit of $x^{+}_\epsilon(t)$ that we obtain for values of $t$ sufficiently large for  
 $x^{+}_\epsilon(t)-x^{-}_\epsilon(t)$ to  remain approximately constant. 
The results are not sensitive to the exact value of $\epsilon$: 
comparison with results obtained for $\epsilon=0.001$, $0.01$ and $0.1$  resulted in less than 1$\%$ of difference in the speed of the front. 
  
 \begin{figure}[!]    
\centerline{(a) Regime I} 
 \centerline{\includegraphics[width=0.85\linewidth]{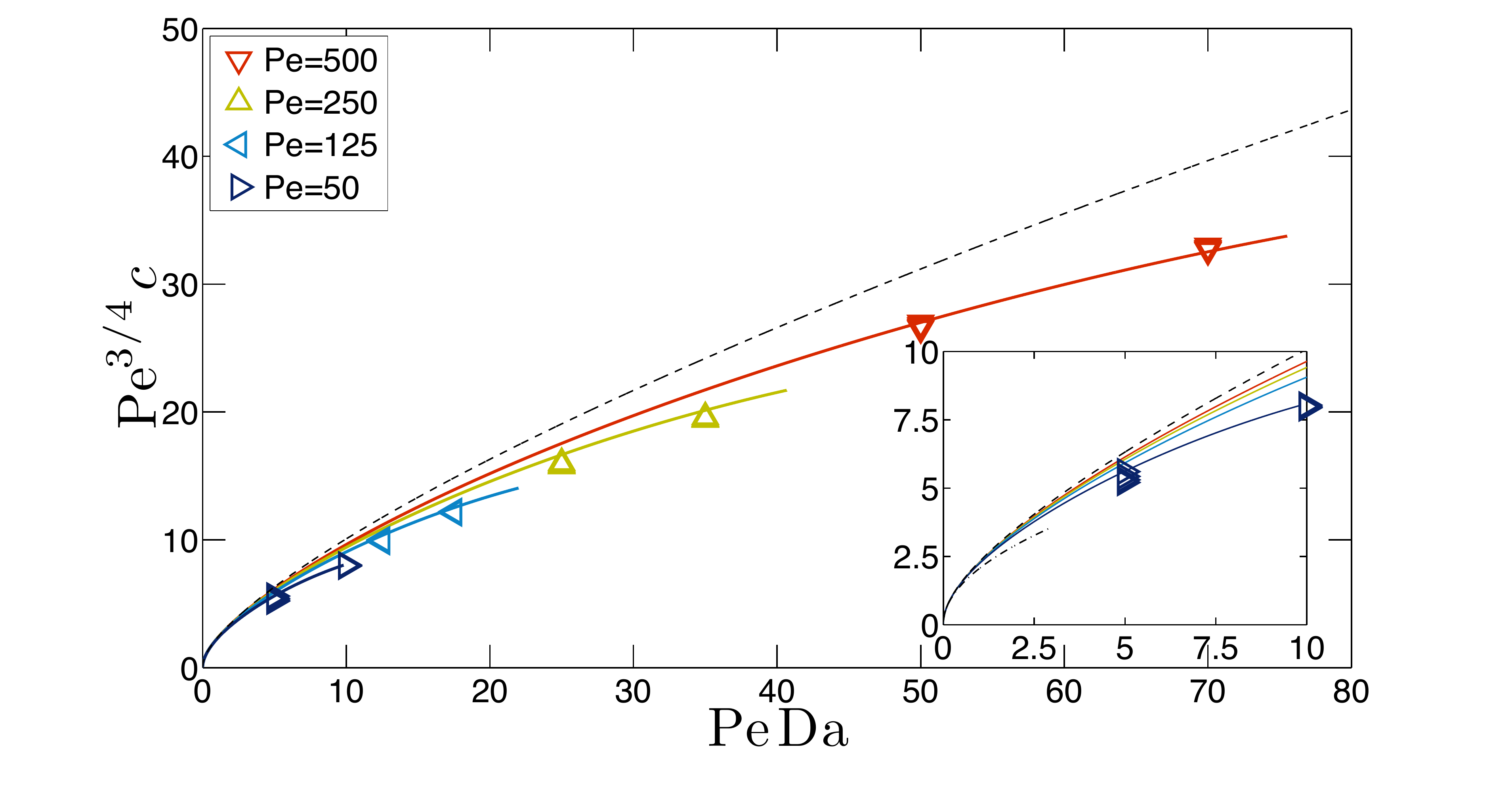}}   
  \centerline{(b) Regime II}  
  \centerline{\includegraphics[width=0.85\linewidth]{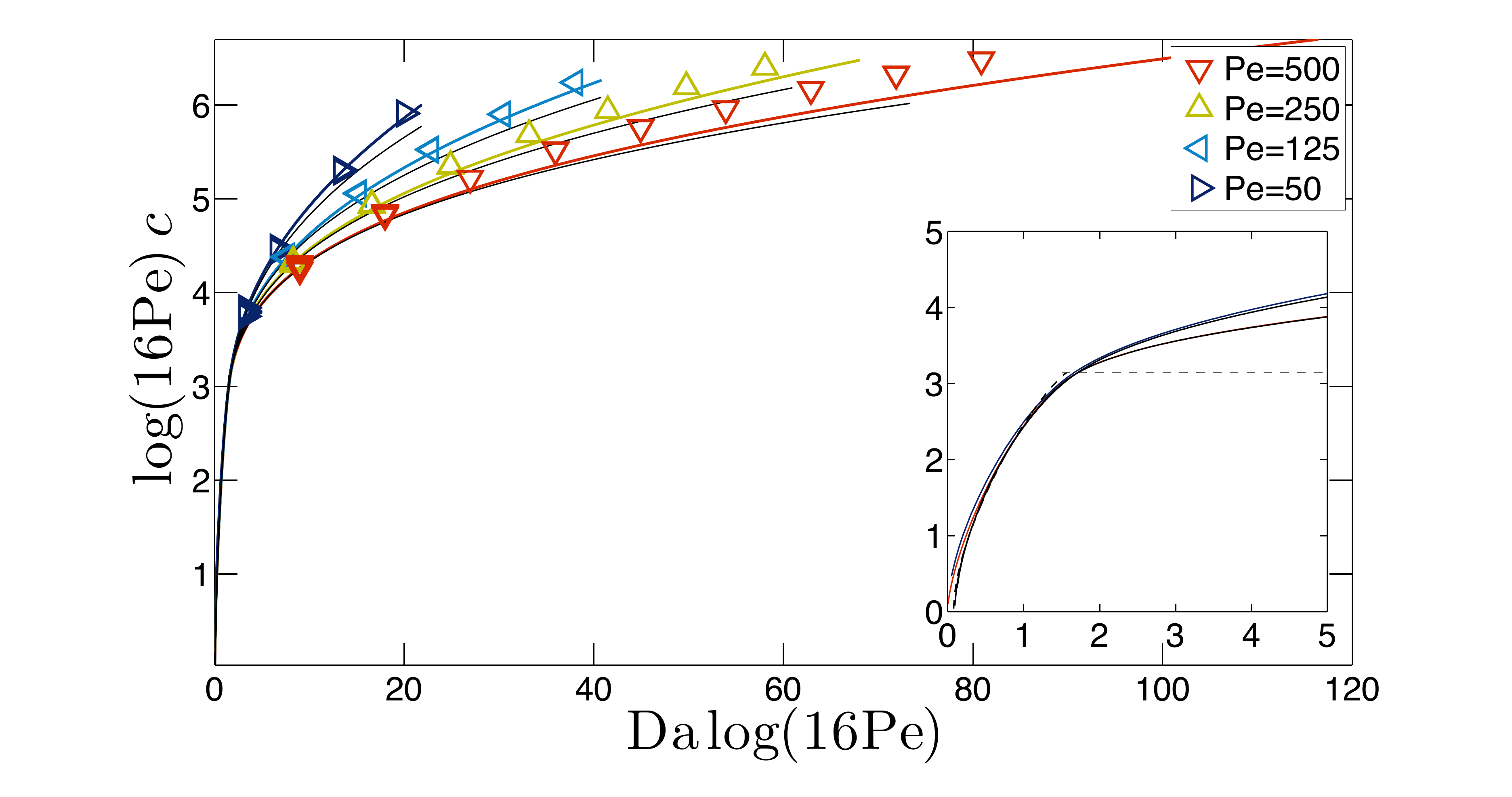}}  
   \centerline{(c) Regime III}         
   \centerline{\includegraphics[width=0.85\linewidth]{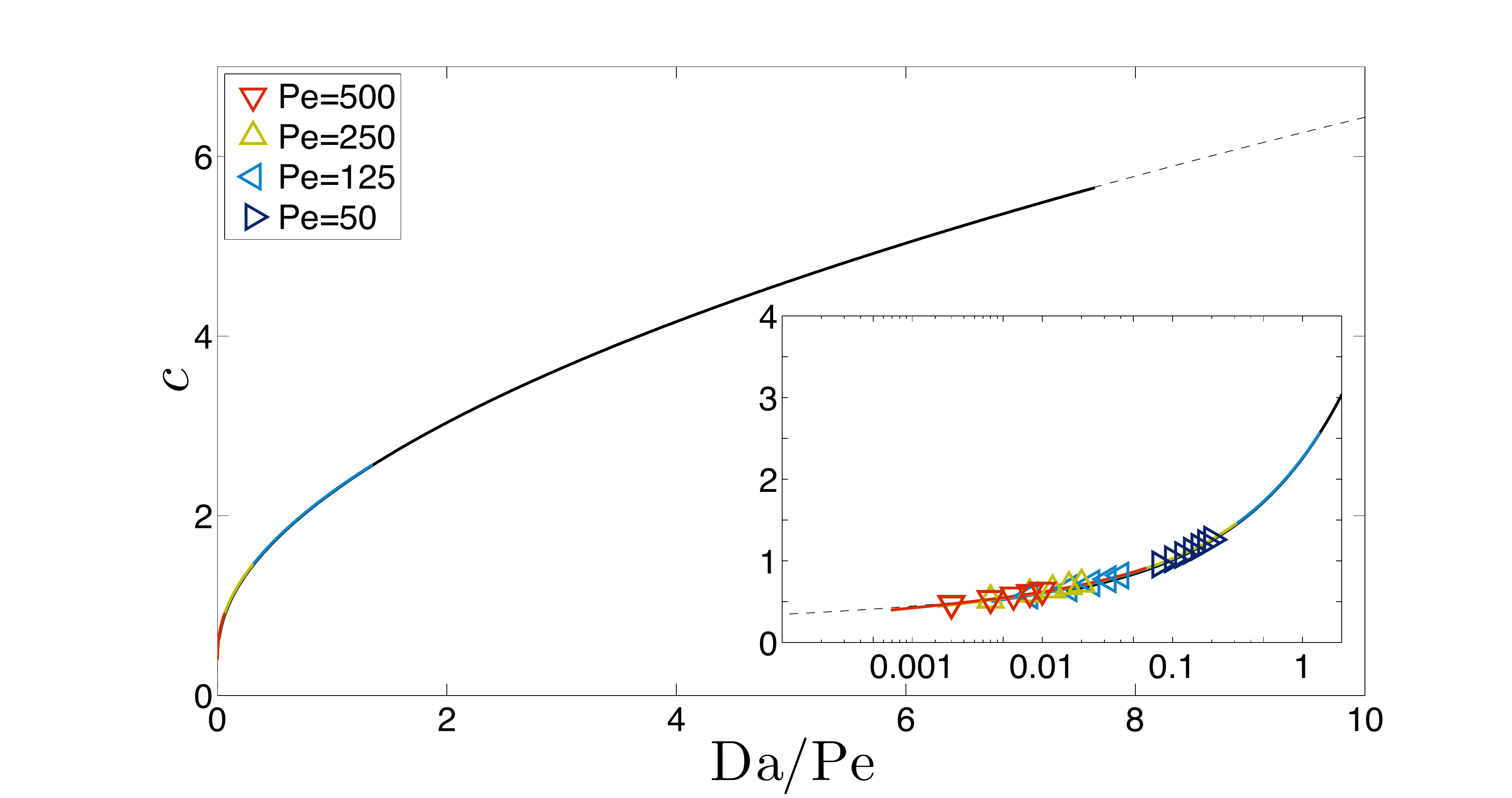}}                 
     \caption{(Color online). Comparison between asymptotic and numerical results of the front speed $c$ in all three regimes and for various values of $\Pe$.
	  Solutions of the eigenvalue problem  are shown in  thick, grey (colored) solid lines. 
	 Results of the full numerical simulations 
	  are shown as symbols.
The dashed thin  lines are the asymptotic predictions.  (a) Regime I showing 
the asymptotic prediction \eref{speed_regI} valid for $\Da=O(\Pe^{-1})$; the diffusive approximation  \eref{speed_regIa} is also shown in the inset that magnifies the small-$\Pe\Da$ region (dashed-dotted line).
(b) Regime II showing the asymptotic prediction \eref{speed_regII} valid for  $\Da=O((\log(16\Pe)^{-1})$;
the corrections to prediction \eref{speed_regII} obtained by iterating
\eref{f0_RegimeIIappcorr} are also shown (thin solid black lines).
The inset focuses  on two values of $\Pe=50,500$ and on a smaller region of values of $\Da\log(16\Pe)$.
(c) Regime III showing the asymptotic prediction \eref{c3} valid for $\Da=O(\Pe)$, with inset focussing on small values of $\Da/\Pe$. 
	 }      
	\label{fig:num} 
\end{figure} 

The two sets of numerical results are shown in Figure \fref{num} along with the corresponding prediction for each regime, respectively derived from \eref{speed_regI}, \eref{speed_regII} and \eref{c3}.
The speeds obtained from the eigenvalue equation \eref{eval2} are in excellent agreement with the corresponding   values obtained from the full numerical simulations of the FKPP equation \eref{FKPP}. 
This is especially the case when $\Da\gtrsim 10 \, \Pe^{-1}$.  
 For $\Da\approx 5 \, \Pe^{-1}$, we observe a small dependence of the speed value on the threshold $\epsilon$ that is used to define the right endpoint of the front (see inset in  Figure \fref{num}(a)). 
This dependence is due to the particularly long integration times and computational domain that are necessary to capture this slowly advancing, wide front (see   Fig. \fref{fronts}(a)). 
As $\Da$ increases to $O(1)$ values and beyond, the solutions to \eref{FKPP} and \eref{eval2} become progressively localized, with the smallest
lengthscales being $O(\delta_2)$, which are challenging  to resolve when $\Pe \gg 1$. 
This is partly reflected in Figure  \fref{num}(b) where for the high values $\Pe=250,500$, 
the agreement between the two sets of numerical results is not as close as for the moderate values $\Pe=50,125$, with the difference increasing with $\Da$.  
In Figure \fref{num}(c), where the speed is unscaled,  the agreement is excellent.  
However, we were not able to obtain sufficiently accurate speed values when $\Da/\Pe=O(1)$ from
either \eref{FKPP} or \eref{eval2} due to the numerical limitations  when $\Da,\,\Pe \gg 1$. 
 
It is clear that in all three regimes,  the asymptotic predictions  become increasingly accurate as the value of $\Pe$ increases. 
In Regime II, the agreement is very good for all values of $\Pe$  when $\Da\log(16\Pe)$ is small. However, when $\Da\log(16\Pe)$ is large,  we need to employ  higher-order corrections to   \eref{speed_regII} (which are obtained via \eref{f0_RegimeIIappcorr}). These capture very well  the slow growth of the speed values when $\Da\log(16\Pe)\gg 1$, particularly so for $\Pe=250$, $\Pe=500$. 
In Regime III, the agreement is excellent for all values of $\Pe$. 

As expected, the asymptotic expressions are valid over a broad range of values of their argument, restricted only by the range of validity of each regime. Taken together, they cover the entire range of $\Da$ to provide convenient approximations for the front speed $c$,
including when $\Pe$ and/or $\Da$ are so large that direct numerical computations are challenging.

\section{Conclusion}\slab{conc}

In this paper, we study the classic problem of FKPP front propagation in a cellular flow. We examine in detail the asymptotic form of the front speed $c$ in the limit of large P\'eclet number $\Pe$  corresponding to a diffusion that is weak compared to advection, and for arbitrary values of the Damk\"ohler number $\Da$, i.e., arbitrary reaction rate. This is achieved by a careful asymptotic analysis of the two-dimensional eigenvalue problem from whose solution $c$ can be deduced. This is complicated by the non-uniformity of the problem: depending on the relation between $\Da$ and $\Pe$, different regimes emerge which require different asymptotic methods and lead to different expressions for $c$. 

Specifically, we identify the three distinguished regimes listed in Table \ref{tbl1}. In each regime, the front speed is given in terms of a transcendental function of a suitable combination of $\Pe$ and $\Da$. Each function is determined by solving a ($\Pe$-independent) one-dimensional problem: an ordinary differential equation in Regime I, an integral eigenvalue problem in Regime II, and an optimisation problem in Regime III. These problems need to be solved numerically in general, though at a much reduced computational cost compared with the original two-dimensional eigenvalue problem thanks to the dimensional reduction, the independence on $\Pe$, and the single scale of the solution. Closed-form expressions are obtained by considering asymptotic limits of the one-dimensional functions characterising Regimes I, II and III, leading to the 
subregimes listed in Table \ref{tbl2}. By verifying that the same expressions for $c$ can be obtained by suitable limits of both Regimes I and  II on the one hand, and of both Regimes II and III on the other, we confirm that our formulas cover the full range of values of $\Da$. We emphasise that the closed-form formulas valid in the various subregimes apply to most of the $(\Pe,\Da)$-plane for $\Pe \gg 1$, with the more complex distinguished expressions only required in the comparatively narrow regions defined by $\Da \,\Pe = O(1)$, $\Da \, \log \Pe = O(1)$ and $\Da/\Pe=O(1)$. 

   Our analysis reveals previously unchartered behaviour. Only two sublimits are intuitively obvious: the first (IIIb, $\Da \gg \Pe$) arises when the reaction is so fast that advection can be  neglected, so that the front speed is the familiar bare speed, dimensionally $c_{\textrm{IIIb}} = c_0 = 2 \sqrt{\kappa/\tau} = 2 U \sqrt{\Da/\Pe}$, obtained in the absence of flow. The other obvious sublimit (Ia, $\Da \ll \Pe^{-1}$) arises when the reaction  
  is slow enough that the front spreads across many flow cells; in this case, homogenisation results which describe the combined effect of advection and diffusion through an effective diffusivity $\kappa_\mathrm{eff}$ apply, and the front speed is estimated by replacing $\kappa$ by $\kappa_\mathrm{eff} = 2 \nu \Pe^{1/2} \kappa$ in the bare speed to obtain $c_{\textrm{Ia}} = \sqrt{8 \nu} U  \Pe^{-1/4} \Da^{1/2}$. These two explicit expressions provide estimates for $c$ for extreme values of $\Pe$, 
but since their  ratio  $c_{\textrm{Ia}}/c_{\textrm{IIIb}} = \sqrt{2 \nu} \Pe^{3/4}$ is asymptotically large, 
they provide little indication (bar a lower bound for $c_{\textrm{IIIb}}$) for the front speed for $\Da$ away from these extremes. 
Our asymptotic results, in contrast, pinpoint the behaviour of $c$. They describe, in particular, the very slow growth of $c$ with $\Da$ in Regime IIb/IIIa where, to the lowest order ignoring logarithmic corrections, $c \sim \pi / \log \Pe$ is independent of $\Da$. This scaling, proposed heuristically in \cite{Abel_etal2002,Cencini_etal2003}, is here derived in two ways, from the integral eigenvalue problem of Regime II and from the optimisation approach of Regime III. It can be traced to the behaviour of fluid-particle motion near the cell corners: the front in this regime is controlled by motion along the separatrix which is fast along most of the separatrix but very slow near the corners since these are   stagnation points. As a result, the motion of particles determining the front is akin to a random walk on the lattice of stagnation points. It is not difficult to show that the relevant waiting time, namely the typical time by diffusion to move particles across the stagnation point scales like $\log \Pe$, thus explaining  the form of $c$. A more complex dependence on $\log \Pe$ holds in the entire Regime II,  reflecting the same physical phenomenon although complicated by a non-trivial behaviour between stagnation points. 

We conclude by noting that most of the rigorous work on the asymptotics of FKPP front speed focuses on a single large parameter, namely the P\'eclet number, assuming either that $\Da=O(\Pe^{-1}) \ll 1$ \cite{Heinze2005,RyzhikZlatos2007,NovikovRyzhik2007,Zlatos2010} or that $\Da=O(\Pe) \gg 1$ \cite{MajdaSouganidis1994,FreidlinSowers1999}. Our analysis and numerical work demonstrates the richness of the problem when the Damk\"ohler number is allowed instead to take a broad range of value. This richness no doubt extends much beyond the specific cellular flow considered in this paper; extensions that demonstrate this for a wide class of flows would be desirable.

\section*{Acknowledgments}
The authors thank P. H. Haynes, G. C. Papanicolaou and A. Pocheau for helpful discussions. This work was supported by  EPSRC (Grant No. EP/I028072/1).

\appendix

\section{Boundary-layer analysis in Regime I}\slab{RegimeIA}
 We establish expression \eref{deriv2}. The derivation is essentially identical to the one  in \cite[Appendix A.2]{HaynesVanneste2014b}  and is detailed here
for completeness. The differences lie in the matching conditions
\eref{MC} but, despite these differences, the derivative \eref{deriv2} of the eigenfunction 
remains unaltered.

Inside the boundary layer, we use the rescaled variables \eref{var} and denote 
solutions  in the $\pm$ half cells by $\Phi^\pm(\sigma,\zeta)$. The alternating symmetry \eref{phi0a} reads
\beq\elab{sym_bl}
\Phi^\pm(\zeta,\sigma)=\Phi^\mp(\zeta,\sigma+4).
\eeq  
This condition, the boundary conditions \eref{phi0b} and continuity across the half-cells imply that
\begin{subequations}\elab{MC}
\begin{align}
\Phi^\pm(0,\sigma)=\Phi^\mp(0,\sigma+4),\ \ & \partial_\zeta\Phi^\pm(0,\sigma)=0   \\
 & \textrm{for} \ \ 0<\sigma<2 \ \ \textrm{and} \ \ 4<\sigma<6, \nonumber \\
\Phi^\pm(0,\sigma)=\Phi^\mp(0,\sigma), \ \ &
\partial_\zeta\Phi^+(0,\sigma)=-\partial_\zeta\Phi^-(0,\sigma) \elab{MC3}  \\ 
& \textrm{for} \ \  2<\sigma<4 \ \ \textrm{and} \  \ 6<\sigma<8. \nonumber 
\end{align}
\end{subequations}

Introducing expansions of the form \eref{exp}   into \eref{eval2} 
 leads to the sequence  
\begin{subequations}\elab{interior_pb}
\begin{align} 
	&
	\left(
	\partial_{\zeta\zeta}^2
	-\partial_\sigma
	\right)\Phi_{0}^\pm=0 \quad\text{at} \ \  O(1), 
	\elab{phi00app}\\ 
	&
	\left(
	\partial_{\zeta\zeta}^2
	-\partial_\sigma
	\right)\Phi_{k}^\pm+
	\frac{u_1 \hat{q}}{\nor{\bu}^2}\,\Phi_{k-1}^\pm=0 \quad\textrm{at} \ \ O(\Pe^{-{k}/{4}}), \ \ \textrm{for} \ \ k=1,\, 2.
	\elab{phi0k}
  \end{align}
\end{subequations}
The only admissible solution to \eref{phi00app} is a constant: $\Phi_{0}^\pm = \Phi_0 =\mathrm{const}$.
Expressing $\nor{\bu}$ in terms of $\sigma$, \eref{phi0k} becomes
\beq\elab{phi0kb}
	\left(
	\partial_{\zeta\zeta}^2
	-\partial_\sigma
	\right)\Phi_{k}^\pm=
	\mp F(\sigma)\,\hat{q}\Phi_{k-1}^\pm, 
\eeq	
where $F(\sigma)=(2\sigma-\sigma^2)^{-1/2}$ for $0<\sigma<2$, $F(\sigma)=0$ for $2<\sigma<4$, and $F(\sigma+4)=-F(\sigma)$.
 It follows that the $k=1$ solution to \eref{phi0kb}   satisfies $\Phi_{1}^\pm(\zeta,\sigma+4)=-\Phi_{1}^\pm(\zeta,\sigma)$ which once combined with 
 \eref{sym_bl} yields $\Phi_{1}^\pm(\zeta,\sigma)=-\Phi_{1}^\mp(\zeta,\sigma)$.
The matching conditions \eref{MC} now become
\begin{subequations}\elab{MC2app}
\begin{align}
&\Phi^\pm_1(0,\sigma)=-\Phi^\mp_1(0,\sigma),\ \ \partial_\zeta\Phi^\pm_1(0,\sigma)=0\ \ \textrm{for} \ \ 0<\sigma<2 \ \ \textrm{and} \ \ 4<\sigma<6, \elab{MC2b}\\
&\Phi^\pm_1(0,\sigma)=0, \ \  
\partial_\zeta\Phi^+_1(0,\sigma)=-\partial_\zeta\Phi^-_1(0,\sigma)
\ \ \textrm{for} \ \ 2<\sigma<4 \ \ \textrm{and} \ \ 6<\sigma<8.\elab{MC3b} 
\end{align}
\end{subequations}
Defining $G(\sigma)$ by $G'(\sigma)=F(\sigma)$ with $\int_0^8 G(\sigma) \, \dd \sigma=0$,
we write the solution to \eref{phi0kb} for $k=1$ in terms of $G$ and $\rho$ as 
\beq\elab{Phi_1}\\  
\Phi_{1}^\pm(\zeta,\sigma)=\pm
\hat{q}\left(G(\sigma)+\rho(\sigma,\zeta)\right) \Phi_0.
 \eeq
Here $\rho(\zeta,\sigma)$  satisfies
$\partial_{\zeta\zeta}^2\rho =\partial_\sigma\rho$ with $\rho\rightarrow 0$ as $\zeta\rightarrow\infty$. The boundary conditions on the cell boundaries impose that  $\partial_\zeta\rho(0,\sigma)=0$ for  $0<\sigma<2$, $4<\sigma<6$ and $\rho(0,\sigma)=-G(\sigma)$ otherwise. 
The problem describing $\rho$ is essentially the same
as the problem solved by \cite{Soward1987} in the Appendix  
(the exact correspondence is achieved upon multiplication of $\rho$ by $-2/\pi$ and its translation so that $\sigma\mapsto\sigma-2$). The key result  is 
\beq\elab{useful}
\frac{2}{\pi}\int_0^\infty\rho(\zeta,0) \, \dd \zeta=-2\nu,
 \eeq
where $\nu$ is the constant defined in \eref{effective}.

An approximation to $\partial_\zeta\Phi_{1}^\pm$ and $\partial_\zeta\Phi_{2}^\pm$ as $\zeta\rightarrow\infty$ is obtained by 
noting that  the leading-order behaviour of $\Phi_{k}$ at large values of $\zeta$
is controlled by its average around the streamline so that
\beq
\Phi_{k}^\pm\sim\av{\Phi_{k}^\pm}\equiv\frac{1}{8}\int_0^8\Phi_{k}^\pm\,\dd\sigma \quad\textrm{for} \ \ k=1, \, 2, \ \ \textrm{as} \ \ \zeta\rightarrow\infty.
\eeq
We integrate  \eref{phi0kb}  first  over $\sigma$, then over $\zeta$
 to obtain that $\lim_{\zeta\rightarrow\infty}\partial_{\zeta}\av{\Phi_{1}^\pm}=0$ and 
\beq
		\lim_{\zeta\rightarrow\infty}\partial_{\zeta}\av{\Phi_{2}^\pm}=
		-\frac{1}{4}\hat{q}^2\int_0^2 \dd\sigma F(\sigma)\int_0^\infty \dd\zeta\rho(\zeta,\sigma)
		=-\frac{\pi}{4}\hat{q}^2\int_0^\infty \dd\zeta\rho(\zeta,0)
		=-\frac{\nu\pi^2}{4}\hat{q}^2,
		\elab{derivzeta2}
\eeq
		where we have combined  \eref{sym_bl} and \eref{MC} to find that $\partial_{\zeta}\av{\Phi_{k}^\pm}=0$ for $\zeta=0$, $k=1,\, 2$ when $0 < \sigma < 2$. This derivation uses that $\partial_\sigma\int_0^\infty \rho(\zeta,\sigma) \, \dd\zeta=-\partial_\zeta\rho(0,\sigma)=0$ for $0<\sigma<2$, that $\int_0^2 F(\sigma) \, \dd \sigma =\pi$, and \eref{useful}.

\section{Eigenvalue problem in  Regime II}\slab{appII}
We derive the explicit form of the asymptotic eigenvalue problem \eref{sys}. 
The functions $\widehat{\Phi}^\pm$, related to the eigenfunction $\Phi_{0}^\pm$ in the `$\pm$' cells via  \eref{amp_hat}, satisfy the heat equation \eref{heat} away from the cell corners. Using this and the no-flux boundary conditions \eref{phi0b} makes it possible to write
\begin{subequations}\elab{corners}
	\begin{align}
	\widehat{\Phi}^+(\zeta,2^-)&=(\H_++\H_-)\widehat{\Phi}^+(\zeta,0^+),\\
	\widehat{\Phi}^+(\zeta,6^-)&=(\H_++\H_-)\widehat{\Phi}^+(\zeta,4^+),
\end{align}
where $\H^\pm$ are the `time-2' heat-flow maps defined in \eref{heatop}. Continuity of $\Phi^+$  across  the half-cells implies that 
\begin{align}
	\widehat{\Phi}^+(\zeta,4^-)&=\H_-\widehat{\Phi}^+(\zeta,2^+)+\H_+\widehat{\Phi}^{\mathrm{R}}(\zeta,2^+),\\
	\widehat{\Phi}^+(\zeta,0^-)&=\H_-\widehat{\Phi}^+(\zeta,6^+)+\H_+\widehat{\Phi}^{\mathrm{L}}(\zeta,6^+),
\end{align}
where $\widehat{\Phi}^{\mathrm{R}}$ and $\widehat{\Phi}^{\mathrm{L}}$ are the  solutions inside the neighbouring `$-$'
cells, located on the left and the right of the `$+$' cell, respectively. Using the  alternating symmetry \eref{phi0a} and the definition \eref{amp_hat} further gives   
\begin{align}
\widehat{\Phi}^{R}(\zeta,2^+)&=e^{-\pi q}\widehat{\Phi}^{+}(\zeta,6^+),\\
 \widehat{\Phi}^{L}(\zeta,2^+)&=e^{\pi q}\widehat{\Phi}^{+}(\zeta,2^+).
\end{align}
\end{subequations}	
Employing the jump condition \eref{jump} to eliminate the upstream functions $\widehat{\Phi}^+$ (at $\sigma=0^-,\, 2^-,\, 4^-$ and $6^-$) in favour of the downstream ones (at  $\sigma=0^+,\, 2^+,\, 4^+$ and $6+$)  reduces \eref{corners} to the eigenvalue problem
\begin{subequations}\elab{integralpbA}
\beq
(16\Pe)^{{f_0}/{2}}\widehat{\bm\Phi}(\zeta)=(\mathbfcal{K}\widehat{\bm\Phi})(\zeta) 
\elab{integraleqnA}
\eeq
where 
\beq
\widehat{\bf\Phi}= 
\begin{pmatrix}
	\widehat{\Phi}^+(\zeta,0^+)\\[2pt]
	\widehat{\Phi}^+(\zeta,2^+)\\[2pt]
	\widehat{\Phi}^+(\zeta,4^+)\\[2pt]
	\widehat{\Phi}^+(\zeta,6^+)
\end{pmatrix},	
\quad 
\mathbfcal{K} = 
 \begin{pmatrix}
  0 & \ee^{\pi q}\L_- & 0 & \L_+\\
  \L_++\L_- & 0 & 0 & 0 \\
  0  & \L_+  & 0 & \ee^{-\pi q}\L_-  \\
  0 & 0 & \L_++\L_-& 0
 \end{pmatrix},
\eeq
\end{subequations}
and $\L_\pm=\zeta^{f_0}\H_\pm$

\section{Matching of Regimes II and III}\slab{f0_asym_large}
In this section, we derive the   form of $f$ in   
subregime IIb  and show that it matches the corresponding expression in   subregime IIIa.  
We seek an approximation to the principal eigenvalue $\lambda$ of $\mathbfcal{K}$ -- or more accurately to $\log \lambda$ from which $f_0$ is inferred -- in the limit $q \gg 1$ and hence $f_0 \gg 1$. It turns out to be advantageous to consider
\beq
	(\mathbfcal{K}^2\widehat{\bm\Phi})(\zeta)
	=\lambda^2\widehat{\bm\Phi}(\zeta),
\eeq
noting the approximations
\beq
\widehat{\bf\Phi}(\zeta)=	\begin{pmatrix}
		\widehat{\Phi}(\zeta,0^+)\\[2pt]
		\widehat{\Phi}(\zeta,2^+)\\[2pt]
		0\\[2pt]
		0
	\end{pmatrix}
	+O(\ee^{-\pi q})
\quad \textrm{and}	\quad
\lambda^2(q,f_0)=\ee^{\pi q}{\Lambda}(f_0)+O(1). \elab{lambdasq}
\eeq
Here, ${\Lambda}(f_0)$   is the principal eigenvalue of the integral equation
\beq
{\Lambda}(f_0)\,\varphi(\zeta)=
(\I_-
\varphi)(\zeta)
+
(\I_+
\varphi)(\zeta), \quad \textrm{with} \ \ \I_\pm= \L_-\L_\pm.  \elab{reduced_int}
\eeq
Note that the above simplification is possible because  $\L_-\L_+$ is the adjoint of $\L_+\L_-$ (and thus they share the same eigenvalues).

The asymptotic behaviour of 
 $\Lambda(f_0)$ for $f_0\gg 1$  is obtained  by introducing the rescaling
 $\zeta=f_0^{1/2}z$ into \eref{reduced_int}.  
It now becomes natural to employ a  WKB expansion for the principal eigenfunction  so that, 
at leading order, $\varphi(z)\sim \exp(f_0 S(z))A(z)$, where $S(z)$ and $A(z)$ remain to be determined.
Thus,  
\beq
(\I_\pm\varphi)\left(z\right)
\sim 
\frac{f_0^{f_0+1}}{8\pi}\int_0^\infty dz_1
\int_0^\infty dz_2
 \exp\left(
 f_0 (h_\pm(z,z_1,z_2)+S(z_2))
 \right)
 A(z_2),
\eeq
where, from the definition of $\H_\pm$  in \eref{heatop},
\beq
h_\pm(z,z_1,z_2)=\log (zz_1)-(z\mp z_1)^2/8-(z_1+z_2)^2/8.
\elab{hpm}
\eeq

The contribution of 
$(\I_-\varphi)(z)$ to  \eref{reduced_int} is subdominant with respect to 
$(\I_+\varphi)(z)$ because, when $f_0\gg 1$, 
 $\exp(f_0h_-)$ is exponentially smaller than  $\exp(f_0h_+)$ 
for all $z_1,z_2>0$. 
Using Laplace's method in  \eref{reduced_int} we obtain that 
\beq
a\equiv\lim_{f_0\rightarrow\infty}f_0^{-1}\log(\Lambda(f_0)f_0^{-f_0}) =h_+(z,z_1,z_2)+S(z_2)-S(z), 
\elab{aas}
\eeq
where $z_1(z)$ and $z_2(z)$ are determined by the saddle-point conditions $\partial_{z_1} h_+=0$ and $\partial_{z_2}h_+=-S'(z_2)$. The constant $a$ governs the  asymptotics of $\log \Lambda(f_0)$ and hence of $\log \lambda$.

Now, the right-hand side of \eref{aas} can be obtained without knowledge of $S(z)$ if it is evaluated at the solution $Z$ of $z_2(Z)=Z$. We now obtain expressions for $z_1(Z)$ and $Z$. Note first that  differentiation of \eref{aas} with respect to $z$ gives $\partial_{z}h_+=S'(z)$. Combining this with the saddle-point conditions leads to 
\beq\elab{usefulrel}
\dpar{h_+}{z_1}(Z,z_1(Z),Z)=0 \quad\text{and}\quad \left( \dpar{h_+}{z}+ \dpar{h_+}{z_2} \right) (Z,z_1(Z),Z)=0.  
 \eeq
Using the explicit form of $h_+$ in \eref{hpm}, these equations are readily solved to find that $Z= z_1(Z)=z_2(Z)=\sqrt{2}$, whence
\beq\elab{Lambda_app}
a=\log 2-1\quad\text{and}\quad \log \Lambda(f_0)\sim f_0 \log (2f_0/\ee). 
 \eeq
 Employing the latter expression into  \eref{lambdasq} provides an expression for $2 \log \lambda$ that  we use inside  \eref{f0_RegimeII} to obtain that $f_0(\log(16\Pe)-a)=\pi q+f_0\log f_0+O(1)$.  
It is now relatively straightforward to deduce that  
\beq\elab{large_f0II}
f_0\sim\frac{-\pi q}{\text{W}_{\text{m}}\left(-\pi q (8 \ee \Pe)^{-1}\right)} \quad\textrm{for} \ \ 1 \ll q\ll \Pe,
\eeq
where $W_\text{m}$ denotes the second real branch of the Lambert W function (see e.g., \cite{NIST:DLMF}). 
The upper bound  in \eref{large_f0II} corresponds to the upper value of $q$ for which $f_0$ remains a non-decreasing function of $q$. 

\bibliographystyle{siam.bst}
\providecommand{\noopsort}[1]{}\providecommand{\singleletter}[1]{#1}%

\end{document}